\pdfoutput=1

\documentclass{aastex631}
\usepackage{epstopdf}
\usepackage{xspace}
\usepackage{graphicx}
\usepackage{amssymb}
\usepackage{amsmath}
\usepackage{natbib}
\usepackage{float}
\usepackage{hyperref}

\shorttitle{Complex Structure of the Eastern Lobe of Pic\,A}
\shortauthors{Thimmappa et al.}

\begin{document}

\title{Complex Structure of the Eastern Lobe of the Pictor\,A Radio Galaxy:\\
	 Spectral Analysis and X-ray/Radio Correlations}

\correspondingauthor{R.~Thimmappa}
\email{rameshan@oa.uj.edu.pl}

\author{R.~Thimmappa}
\affiliation{Astronomical Observatory of the Jagiellonian University, ul. Orla 171, 30-244 Krak\'ow, Poland}

\author{\L .~Stawarz}
\affiliation{Astronomical Observatory of the Jagiellonian University, ul. Orla 171, 30-244 Krak\'ow, Poland}

\author{U.~Pajdosz-\'Smierciak}
\affiliation{Astronomical Observatory of the Jagiellonian University, ul. Orla 171, 30-244 Krak\'ow, Poland}

\author{K.~Balasubramaniam}
\affiliation{Astronomical Observatory of the Jagiellonian University, ul. Orla 171, 30-244 Krak\'ow, Poland}

\author{V.~Marchenko}
\affiliation{Astronomical Observatory of the Jagiellonian University, ul. Orla 171, 30-244 Krak\'ow, Poland}

\begin{abstract}
Here we present detailed analysis of the distinct X-ray emission features present within the Eastern radio lobe of the Pictor\,A galaxy, around the jet termination region, utilising the data obtained from the {\it Chandra} X-ray Observatory. Various emission features have been selected for the study based on their enhanced X-ray surface brightness, including five sources that appear point-like, as well as three extended regions, one characterised by a filamentary morphology. For those, we perform a basic spectral analysis within the 0.5--7\,keV range. We also investigate various correlations between the X-ray emission features and the non-thermal radio emission, utilising the high-resolution radio maps from the Very Large Array at GHz frequencies. The main novel findings following from our analysis, regard the newly recognized bright X-ray filament located upstream of the jet termination region, extending for at least thirty kiloparsec (projected), and inclined with respect to the jet axis. For this feature, we observe a clear anti-correlation between the X-ray surface brightness and the polarized radio intensity, as well as a decrease in the radio rotation measure with respect to the surroundings. We speculate on the nature of the filament, in particular addressing a possibility that it is related to the presence of a hot X-ray emitting thermal gas, only partly mixed with the non-thermal radio/X-ray emitting electrons within the lobe, combined with the reversals in the lobe's net magnetic field.
\end{abstract}


\section{Introduction}
\label{sec:Eintro}

Pictor\,A, classified as a Broad-Line Radio Galaxy (BLRG) with the ``classical double'' (Fanaroff-Riley type II) large-scale radio morphology \citep{Simkin99}, and located at the redshift $z=0.035$ \citep{Eracleous04}, is one of the most prominent radio galaxies in the sky, that has become the prime target for detailed multiwavelength investigations in the recent decades, from radio to the X-ray ranges. More recently, it has been also confirmed as a source of high-energy $\gamma$-rays in the {\it Fermi} Large Area Telescope (LAT) all-sky survey \citep{Kataoka11,Brown12,Ackermann15}.

The large-scale radio/X-ray jet in Pictor\,A originates in the galaxy nucleus, and extends up to hundreds of kiloparsecs beyond the host galaxy to the West \citep{Perley97}; the counter-jet is not prominent at radio frequencies, but can be spotted in deep X-ray maps by the {\it Chandra} X-ray Observatory \citep{Hardcastle05}. The hotspots located at both sides of the core at the lobes' edges, mark the termination points of the jet (to the West) and of the counter-jet (to the East); the bright Western hotspot is clearly detected and even, in some cases, resolved at radio, infrared, optical, and X-ray frequencies \citep{Roeser87,Thomson95,Perley97,Wilson01,Werner12,Isobe17,Thimmappa20}. The radio lobes appear in X-rays as a low-surface brightness cocoon surrounding the large-scale jets \citep{Grandi03,Hardcastle05,Migliori07,Hardcastle16}. 

\begin{table}[t!]
	\caption{{\it Chandra} observations used in our analysis of the E lobe in Pictor\,A}
	\label{Tab:EObsID}
\begin{center}
\begin{tabular}{ c c c c c }
		\hline
		\hline
		ObsID & Date (YYYY-MM-DD) & MJD & Exposure [ks] & Detector \\
		\hline
		346   &2000-01-18 & 51561& 	25.8 &ACIS-23678 \\
		12039 &2009-12-07 & 55172&	23.7 &ACIS-235678\\
		12040 &2009-12-09 & 55174&	17.3 &ACIS-235678\\
		11586 &2009-12-12 & 55177&	14.3 &ACIS-235678\\
		14357 &2012-06-17 & 56095&	49.3 &ACIS-235678\\
		14222 &2014-01-17 & 56674&	45.4 &ACIS-235678\\
		16478 &2015-01-09 & 57031&	26.8 &ACIS-235678\\
		17574 &2015-01-10 & 57032& 	18.6 &ACIS-235678\\
		\hline
	\end{tabular}
\end{center}
\end{table}

Extended lobes in radio galaxies and radio quasars, formed as backflows when the jet plasma passes through the termination shock and is turned away at the contact discontinuity between the shocked outflow and the shocked ambient (intergalactic) medium, are particularly prominent at radio frequencies, due to the synchrotron emission of ultra-relativistic electrons. Detailed radio studies of the lobes with the arcsecond angular resolution, often reveal a complex morphology with filamentary structures and tangled polarization patterns \citep[e.g.,][]{Carilli96,Perley97,Feain11,Anderson18}. The X-ray observations of the lobes, carried out with high-angular resolution modern instruments such as {\it Chandra} or XMM-{\it Newton}, allowed to resolve the lobes and to detect the emission consistent with a non-thermal power-law continuum; if due to the invserse-Comptonization of the Cosmic Microwave Background (CMB) radiation by the lobes' electrons, as typically assumed, this gives the volume-averaged lobes' magnetic field intensities at the level of the equipartition values, namely $B \sim 1-10$\,$\mu$G \citep[see][and references therein]{Kataoka05,Croston05}. During the last decades, some of the most prominent and extended lobes in nearby radio galaxies have been also resolved in high-energy $\gamma$-rays by the {\it Fermi}-LAT \citep{Abdo10,Katsuta13,Ackermann16}.

Lobes are expected to be extremely low-density but high-pressure envelops surrounding (and confining) the jets. As ``calorimeters'' for the jet power deposited over the source lifetime, they are believed to be filled solely by ultra-relativistic electrons and magnetic field, with the total internal energy equal to that of the jet bulk kinetic energy  \citep[e.g.,][]{Begelman89}. However, several observational findings have recently been reported on large amounts of a thermal gas within the lobes, providing a prominent contribution to the X-ray radiative output and the pressure balance of the systems \citep{Stawarz13,OSullivan13,Seta13,Wykes13}.

In this paper, we analyze the archival {\it Chandra} data for the extended lobes in Pictor\,A, focusing in particular on the Eastern (E) lobe and the complex E hotspot region; the bulk of the analysis is based on the single pointing ObsID\,14357, with the 49\,ksec exposure \citep[see][for a summary of all the available {\it Chandra} data for Pictor\,A galaxy]{Hardcastle16}. The X-ray maps of the target are compared in detail with various radio maps obtained by \cite{Perley97} with the NRAO\footnote{The National Radio Astronomy Observatory is a facility of the National Science Foundation operated under cooperative agreement by Associated Universities, Inc.} Very Large Array (VLA).\footnote{The authors thank R. Perley for kindly providing all the radio maps which are presented and analyzed in this paper.} 

Throughout the paper we assume the $\Lambda$CDM cosmology with $H_{0} = 70$\,km\,s$^{-1}$, $\Omega_{\rm m} = 0.3$, and $\Omega_{\Lambda} = 0.7$, for which the luminosity distance to the source is 154\,Mpc, and the conversion angular scale is 0.697\,kpc\,arcsec$^{-1}$. The X-ray spectral fits all take into account Galactic absorption, with the corresponding hydrogen column density $4.12 \times 10^{20}$\,cm$^{-2}$ . The photon index $\Gamma$ is defined in the paper as $F_{\varepsilon} \propto \varepsilon^{-\Gamma}$ for the photon flux spectral density $F_{\varepsilon}$ and the photon energy $\varepsilon$.

\begin{figure}[ht!]
	\centering
	\includegraphics[width=0.65\textwidth]{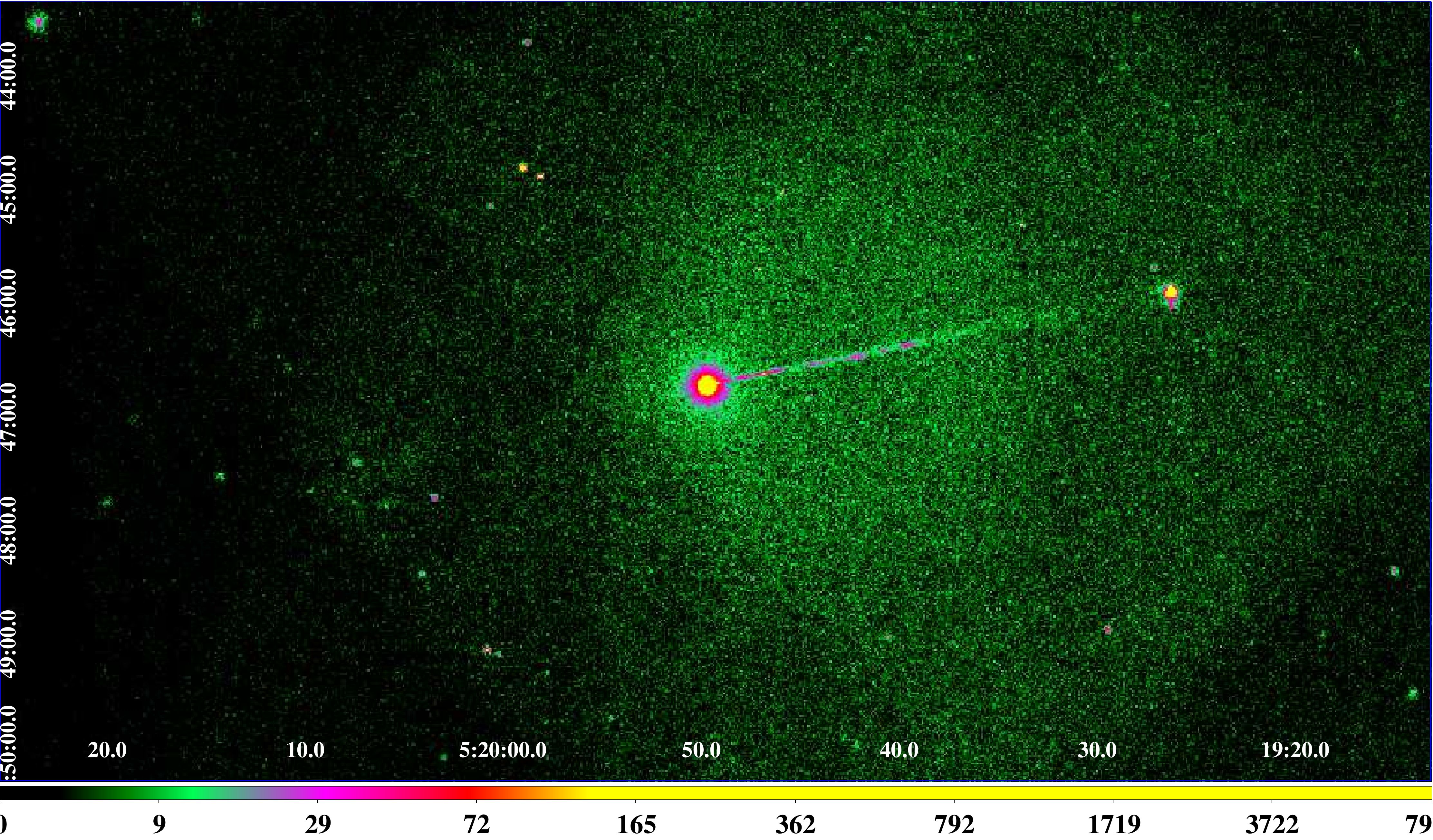}
	\includegraphics[width=0.65\textwidth]{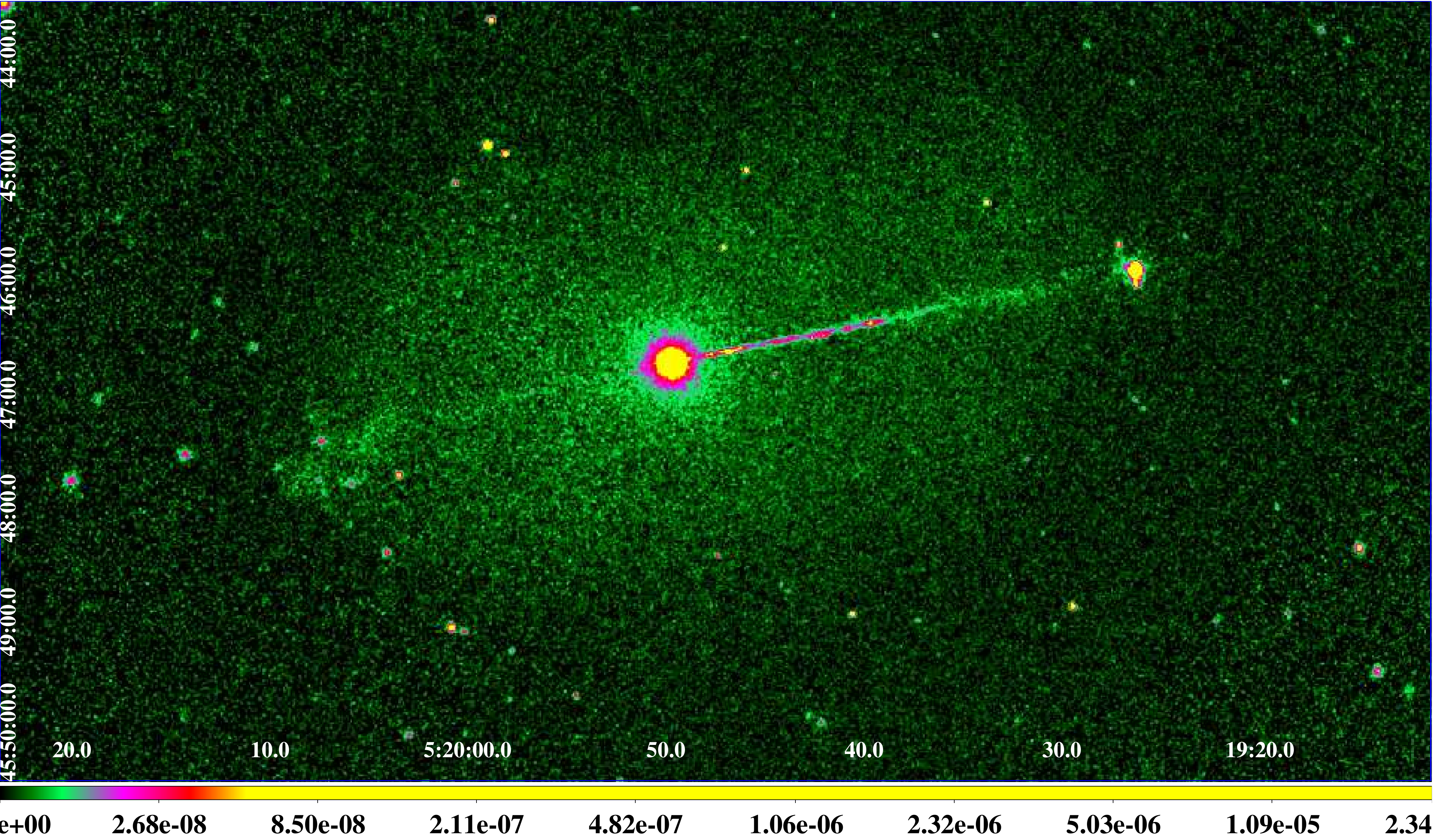}
	\includegraphics[width=0.65\textwidth]{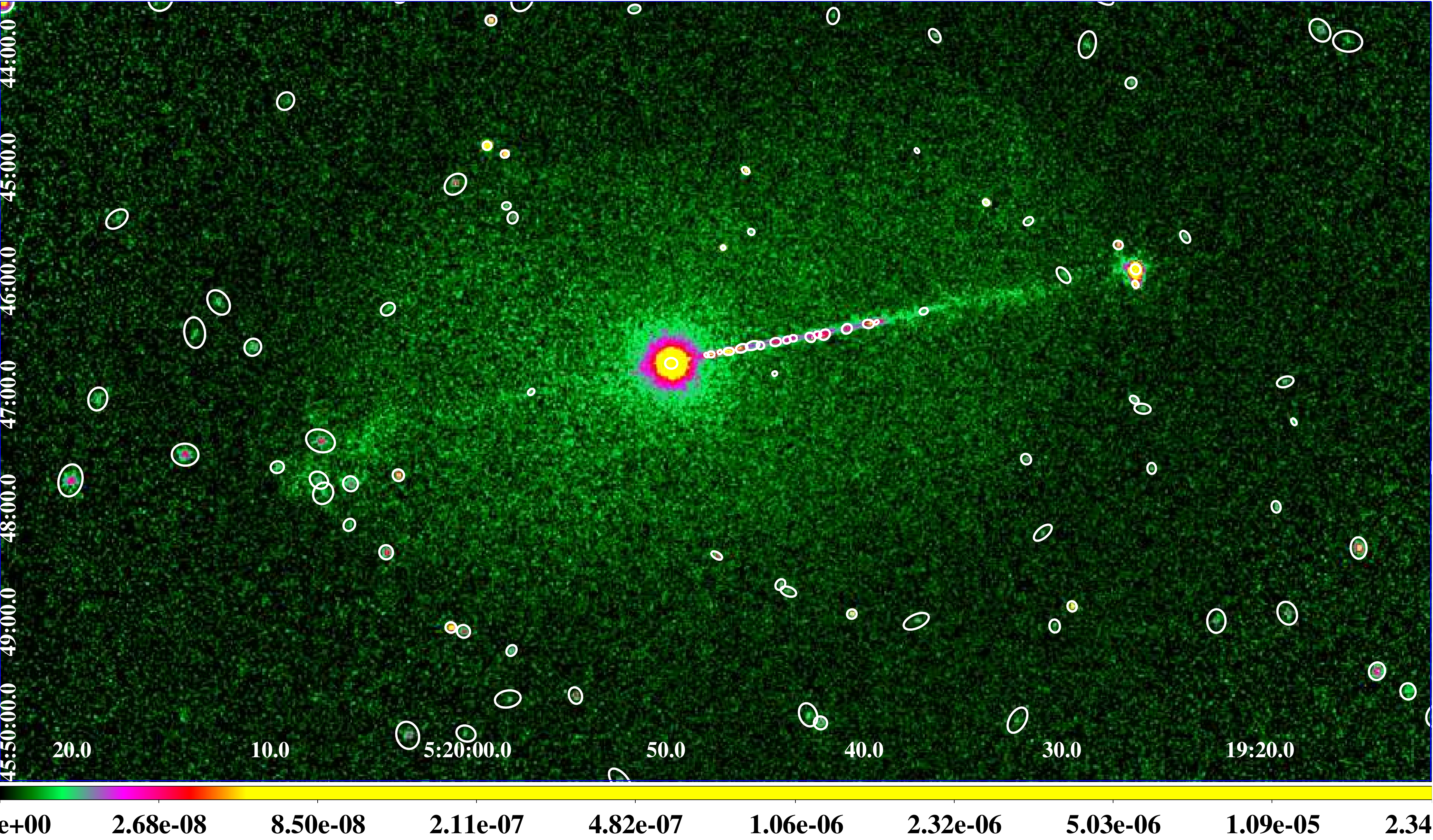}
	\caption{\textbf{Top:} Merged counts image of the Pictor\,A radio galaxy, for the selected eight  {\it Chandra} observations listed in Table\,\ref{Tab:EObsID}, in the energy range 0.5--7.0\,keV, with native $0.492\arcsec$ pixels. Note a much reduced exposure toward the E lobe when compared to the W lobe. \textbf{Middle:} Exposure-corrected merged {\it Chandra} image, smoothed with $3\sigma$ Gaussian radius, revealing the bright core, the jet extending to the North-West from the core, the W hotspot, the weak counterjet to the South-East, the E hotspot region, and the surrounding diffuse lobes. \textbf{Bottom:} Same as in the Middle panel, but with the point and compact sources (denoted by white contours) detected with the {\fontfamily{qcr}\selectfont wavdetect} tool using the minimum PSF method; different sizes of the point/compact sources across the field, reflect the varying PSF and/or sources extension.}
	\label{Fig:Emerge}
\end{figure}

\section{X-ray and Radio Data}

\subsection{Chandra Observations and Data Processing}
\label{sec:EChandra}

Pictor\,A was observed by the {\it Chandra} X-ray Observatory \citep{Weisskopf02} using the Advanced CCD Imaging Spectrometer (ACIS) \citep{Garmire03} on 14 different occasions from January 2000 till January 2015, with the total exposure of about 464\,ksec. Various pointings were optimized for different regions in the extended system, including the nucleus, the Western (W) hotspot, the jet, and the E lobe \citep{Wilson01,Hardcastle05,Hardcastle16}. For our study, we selected a subset of the eight {\it Chandra} pointings, for which the edge of the E lobe --- in particular the E hotspot --- is located away from the chip gap on the detector. In Table\,\ref{Tab:EObsID}, we list the corresponding ObsIDs, observation dates, and exposure times, noting that from those, only in the case of the ObsID\,14357 is the E lobe placed on the back illuminated ACIS-S3 chip, which is characterized by a higher sensitivity at low energies, and a more spatially-uniform, significantly better energy resolution than the front-illuminated chips. In the other ACIS pointings not included in our analysis (ObsIDs 3090, 4369, 14221, 15580, 15593 and 14223), the E hotspot region was located between the ACIS-I or ACIS-S array gaps, in particular between the chips S2, S3 \& I3.

The selected {\it Chandra} observations listed in Table\,\ref{Tab:EObsID}, were reprocessed according to the standard procedure with the CIAO-4.10 package (using the {\fontfamily{qcr}\selectfont chandra\_repro} script) and the Calibration database CALDB-4.7.8 \citep{Fruscione06}, recommended by the CIAO analysis threads. Pixel randomization was removed during the reprocessing, and the readout streaks were removed for each observational data. The {\fontfamily{qcr}\selectfont merge\_obs} script was used to create merged event files for the eight selected pointings; because of the miss-alignments of the source regions in the ACIS-S chip, during the merging process the ObsID\,14357 was taken as the reference observation, due to its high count rates. After merging, the total effective exposure corresponds to 221\,ks. The exposure-corrected images were next generated within the energy band 0.5--7.0\,keV. 
 
\begin{figure}[ht!]
	\centering
	\includegraphics[trim=50 400 20 40, clip, width=0.85\textwidth]{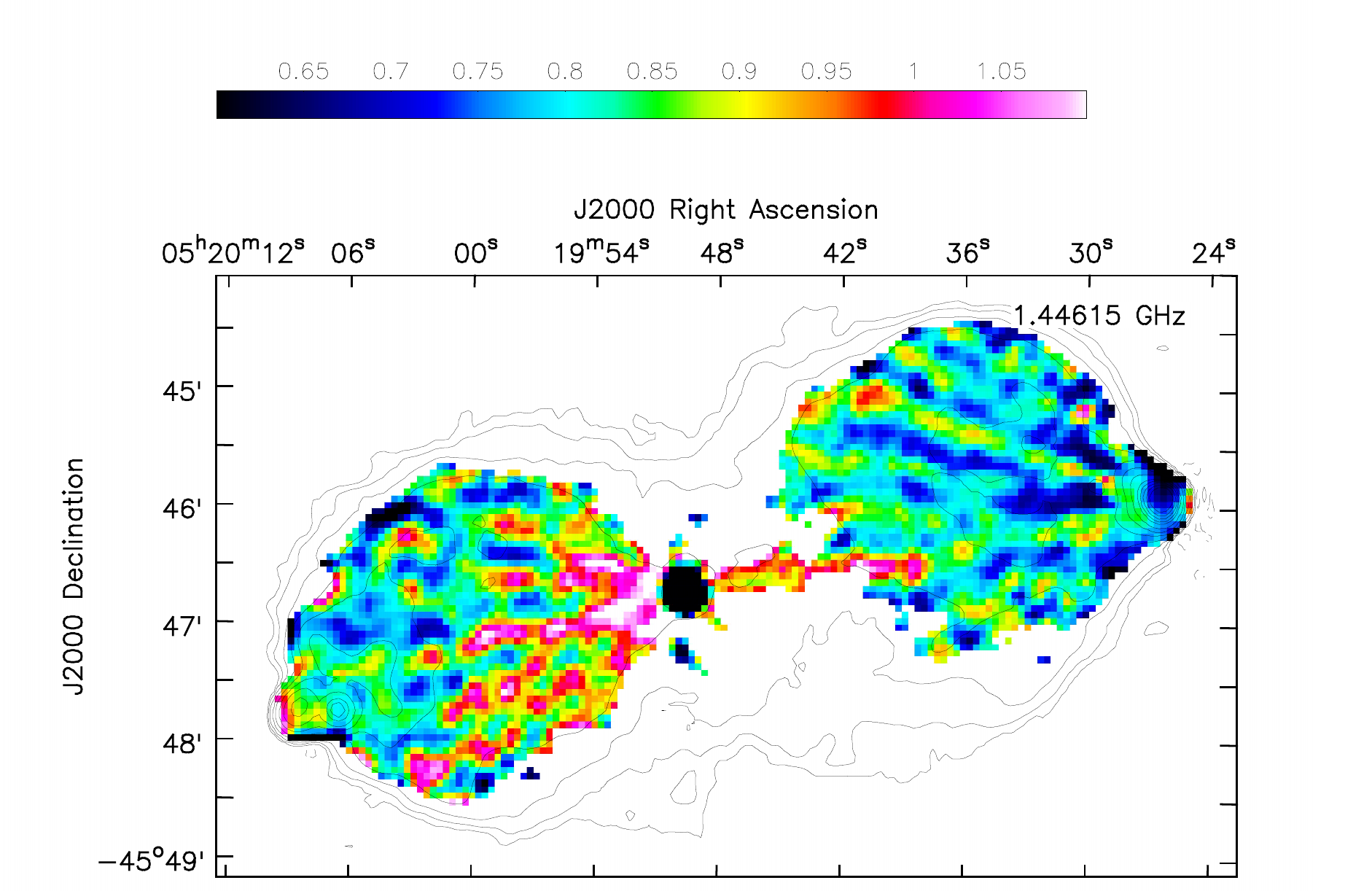}
	\includegraphics[trim=50 400 20 20, clip, width=0.85\textwidth]{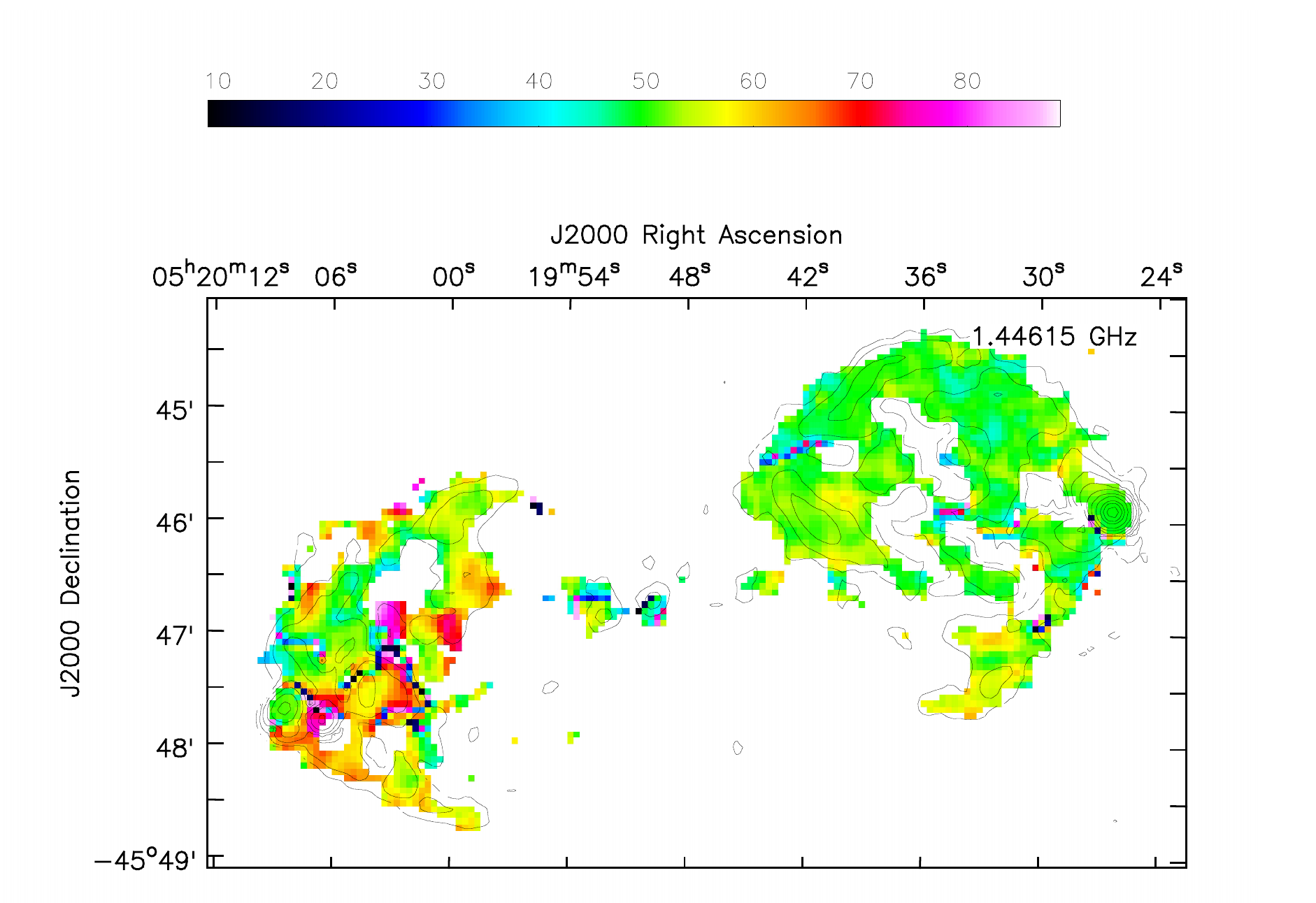}
	\caption{\textbf{Top:} The VLA spectral index map of Pictor\,A radio galaxy, taken between L- and C-band images, with the \emph{total intensity} L-band (1.45\,GHz) contours superimposed, at $10\arcsec$ resolution. Contours started from $3\sigma$ confidence level and are scaled by $\sqrt{2}$. \textbf{Bottom:} The VLA rotation measure (RM) map taken between L- and C-band data, with the \emph{polarization intensity} L-band contours superimposed, at $10\arcsec$ resolution. Contours started from $3\sigma$ confidence level and are scaled by $\sqrt{2}$.}
	\label{Fig:Eradio}
\end{figure}

The merged counts image and the corresponding exposure-corrected image for the entire Pictor\,A radio galaxy, are shown in Figure\,\ref{Fig:Emerge}. On the exposure-corrected map, we searched for point and compact sources with the {\fontfamily{qcr}\selectfont wavdetect} tool implemented in CIAO package, that uses Poisson statistics as appropriate in a low-count regime. The results are shown in the bottom panel of the Figure. Note that, with the applied ``minimum PSF'' method, as recommended when dealing with merged pointings with different off-axis angles, the {\fontfamily{qcr}\selectfont wavdetect} size should match closely the size of sources being detected; for a point source, the corresponding size should therefore reflect the effective Point Spread Function (PSF) at a given position, which may vary across the field. As follows, the algorithm detects the core, a series of knots along the jet, the W hotspot, a net of sources in the E hotspot region, and multiple unrelated background/foreground sources (located mostly outside of the lobes).

For the spectral analysis, we have selected various segments of the E lobe around the counter-jet termination region, based on their enhanced X-ray surface brightness, including five sources that appear point-like, as well as three extended regions, one characterised by a filamentary morphology (see Section\,\ref{sec:Eresults} below). Only one ObsID\,14357 was utilised for the extraction of the source spectra, and for the following spectral modelling. The source spectra were extracted using the {\fontfamily{qcr}\selectfont specextract} script separately for each region in the 0.5--7.0\,keV range, and fitted \emph{simultaneously} with the {\fontfamily{qcr}\selectfont SHERPA} package \citep{Freeman01}. We also attempted a combined spectral analysis involving all the ACIS pointings listed in\,Table\,\ref{Tab:EObsID}; however, due to the fact that in all the observations but ObsID\,14357, the E lobe was located on the front-illuminated chip ACIS-S2 with typically short exposures, this did not provide any significant improvement in terms of the fitting statistics, or the resulting constraints on the model parameters, with respect to the \emph{simultaneous} fitting of the source regions' spectra (together with the background) for the 49\,ks-long ObsID\,14357 observation alone.

\subsection{VLA Observations and Radio Maps}
\label{sec:EVLA}

Detailed VLA studies of Pictor\,A have been presented and discussed extensively in \cite{Perley97}. For the E lobe, the authors noted a higher rms scatter in the derived rotation measure (RM) when compared with the W lobe, though with a similar mean RM value $\sim 45$\,rad\,m$^{-2}$. The polarization degree for both lobes is relatively high $\sim 10-20\%$, increasing up to even $\sim 60-70\%$ at the lobes' edges and at the position of the hotspots. The lobes' spectral index between 0.33\,GHz and 1.45\,GHz is on average 0.8, with some pronounced variations and generally larger values within the E lobe.

In the upper panel of Figure\,\ref{Fig:Eradio}, we show the VLA total intensity L-band contours of Pictor\,A, superimposed on the spectral index map between L- and C-band, with $10\arcsec$ resolution (throughout the paper we use the convention $S_{\nu} \propto \nu^{-\alpha}$, where $S_{\nu}$ is the flux density, $\nu$ is the observed frequency and $\alpha$ is the spectral index). In the lower panel of the Figure, we present the polarized intensity L-band contours of the source, superimposed on the distribution of RM taken between L- and C-band data, with $10\arcsec$ resolution. All the radio images presented here, were made based on the original data analysis by \citet{Perley97}; likewise the radio flux profiles discussed in Section\,\ref{sec:Esurface} below, were calculated  based on VLA maps provided directly by R.~Perley. As shown, the structure of the E lobe around the jet termination region, and in particular upstream of the E hotspot, appear particularly complex on the polarized intensity and RM maps, and this complexity is not obviously reflected in the total intensity, or the spectral index maps. Below we argue that there indeed is a correspondence between the polarization radio maps, and the X-ray {\it Chandra} map of this region.

\begin{figure}[th!]
	\centering
	\includegraphics[trim=57 400 20 30, clip, width=0.86\textwidth]{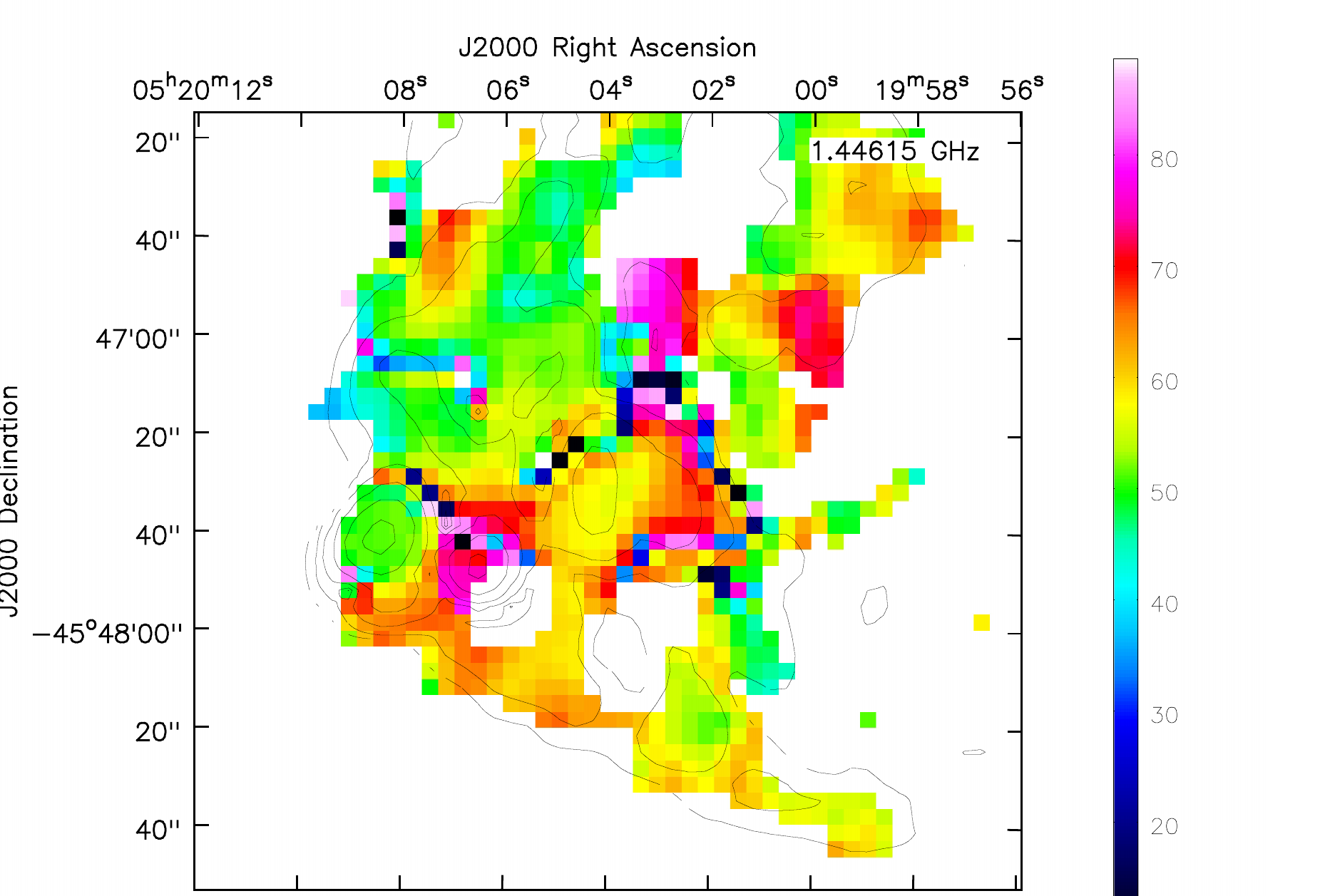}
	\includegraphics[width=0.62\textwidth]{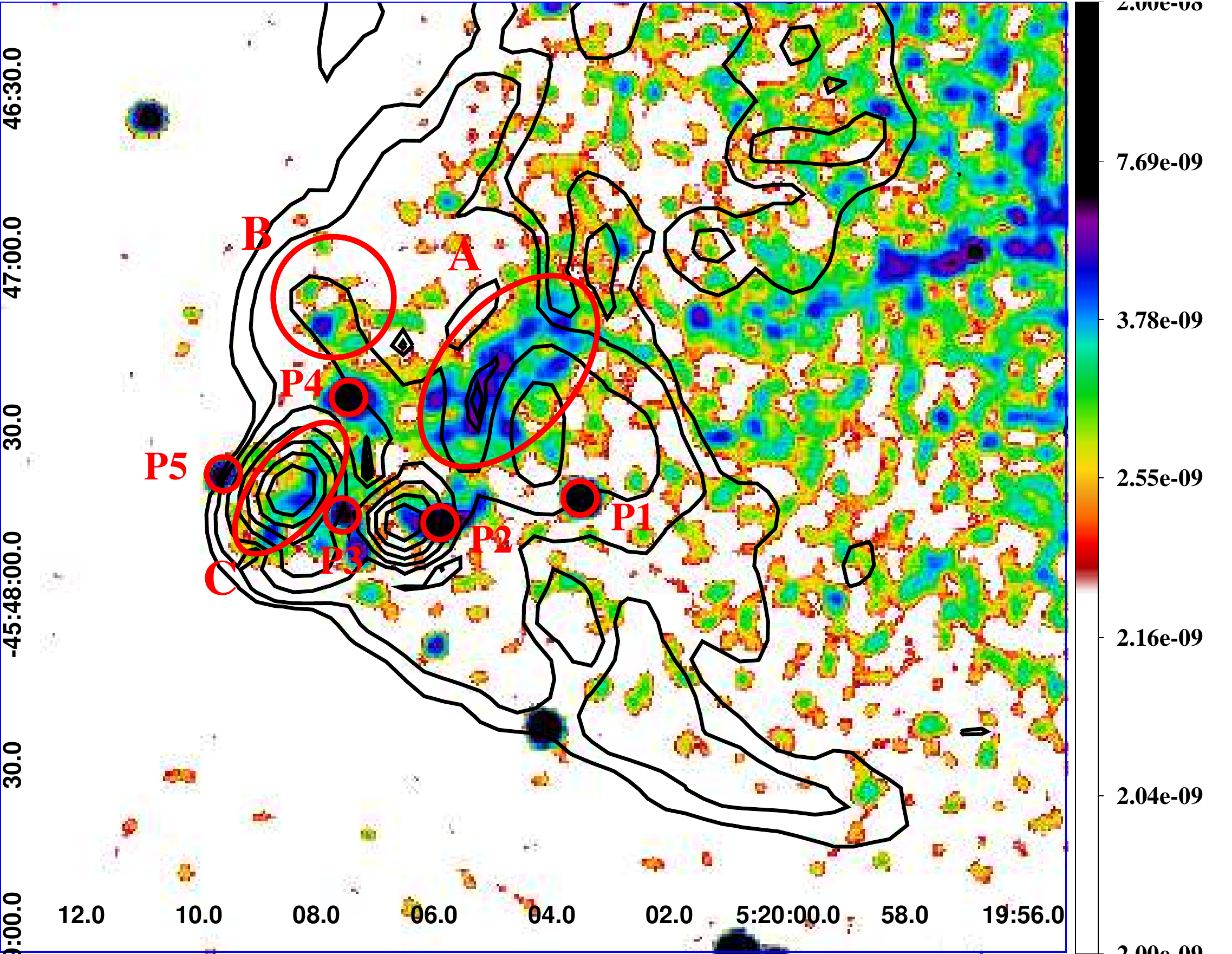}
	\caption{\textbf{Top:} A zoomed view of the RM distribution within the E hotspot region in Pictor\,A, with the polarized intensity L band contours superimposed, at $10\arcsec$ resolution. Contours start from $3\sigma$ confidence level and are scaled by $\sqrt{2}$. \textbf{Bottom:} A zoomed view of the 0.5--7.0\,keV emission of the E hotspot region in Pictor\,A, with the 1.45\,GHz polarized intensity contours (black) superimposed. The 0.5--7.0\,keV {\it Chandra} image is smoothed with $3\sigma$ Gaussian (radius 5\,px). Radio contours start from $3\sigma$ confidence level. Regions selected for the {\it Chandra} data analysis are labeled and indicated by red contours.}
	\label{Fig:EXPol}
\end{figure}

\begin{table}[th!]
	\caption{Power-law fitting results for the selected regions within the E lobe}
	\label{Tab:EPL}
	\begin{center}
		\begin{tabular}{lllll}
			\hline
			\hline
			Region & Size$^{\dagger}$ & Photon index $\Gamma$ & Energy flux $F_{\rm 0.5-7.0 \,keV}$ & Counts$^{\ddagger}$\\
      	    &  [px] & & [10$^{-15}$\,erg\,cm$^{-2}$\,s$^{-1}$] &  \\
      	    
		\hline
		src A & 40/24 & ${1.70}^{+0.23}_{-0.21}$ & ${21.19}^{+3.25}_{- 5.33}$&$ 219$\\
		
		src B & 22 & ${1.89}^{+0.55}_{-0.46}$ & ${4.86}^{+2.84}_{- 1.34}$&$ 68$\\
		
		src C (R2)$^{\star}$ & 28/13 & ${2.17}^{+0.62}_{-0.53}$ & ${5.59}^{+1.94}_{- 2.22}$&$ 66$\\
		
		src P1 (X1)$^{\star}$ & 6 & ${2.27}^{+0.37}_{-0.34}$ & ${5.07}^{+1.94}_{- 0.13}$&$ 41$\\
		
		src P2 (X2)$^{\star}$ &  6 & ${2.15}^{+0.42}_{-0.39}$ & ${3.56}^{+0.19}_{-1.70}$&$ 27$\\
		
		src P3&  6 & ${0.43}^{+0.71}_{-0.74}$ & ${4.55}^{+0.06}_{-2.99}$&$ 12$\\
		
		src P4 (X3)$^{\star}$ &  6 & ${1.13}^{+0.31}_{-0.30}$ & ${7.85}^{+1.11}_{-1.41}$&$ 38$\\
		
		src P5&  6 & ${1.02}^{+0.58}_{-0.57}$ & ${ 3.11}^{+0.17}_{- 1.53}$&$ 12$\\
		
		Background & -- & ${0.27}^{+0.03}_{-0.03}$ & -- & --\\
		\hline
	\end{tabular}
\end{center}
\tablecomments{$^{\dagger}$ Radius in the case of a circle, and major/minor semi-axes in the case of  an ellipse; note the conversion scale 0.492\arcsec/px;\\
$^{\ddagger}$ total number of counts within the $0.5-7.0$\,keV range;\\ 
$^{\star}$ the corresponding/overlaping regions in \citet{Hardcastle16}.}
\end{table}

\section{Analysis Results}
\label{sec:Eresults}

In Figure\,\ref{Fig:EXPol}, we present a zoomed view of the RM distribution within the E hotspot region in Pictor\,A, with the 1.45\,GHz polarized intensity contours superimposed (top panel), as well as of the corresponding 0.5--7.0\,keV {\it Chandra} image, with the 1.45\,GHz polarized intensity contours superimposed (note that in the bottom panel, radio contours start  for clarity only from $3\sigma$ confidence level). The images reveal several interesting features we emphasize below:
\begin{itemize}
\setlength\itemsep{0.0em}
\item The double structure of the hotspot --- which is in fact often observed in classical doubles \citep[see, e.g.,][]{Carilli96} --- is prominent in both total radio intensity and polarized radio intensity maps. The so-called `secondary' hotspot, i.e. the most prominent and outermost radio feature to the East, coincides with some enhancement in the diffuse X-ray emission, but nonetheless appears dramatically weaker at keV photon energies than the W hotspot on the other side of the nucleus; this secondary hotspot is selected for our {\it Chandra} spectral analysis as the extended region ``C'' \citep[overlapping with the region R2 in][]{Hardcastle16}. We do not analyze the spectrum of the `primary' hotspot, i.e. of the more compact radio feature located upstream (to West) from the secondary \citep[see region R1 in][]{Hardcastle16}. Note different values of the RM for both features, namely $\sim 50$\,rad\,m$^{-2}$ and $\sim 80$\,rad\,m$^{-2}$ for the secondary and primary, respectively.

\item There are several bright point/compact X-ray sources in the closest vicinity of the double E hotspot (cf. the bottom panel in Figure\,\ref{Fig:Emerge}), none of which coincides however with the peaks of either total or polarized radio intensity. For our spectral analysis, we select four such distinct regions, labelled as ``P2--P5'' \citep[the three of which overlap with bright compact X-ray sources X1, X2, and X3 in][see Table\,\ref{Tab:EPL}]{Hardcastle16}. On the polarized radio intensity maps, all of these happen to be located almost exactly at the edges of the hotspot's double structure (corresponding to the $3\sigma$ confidence level contours). We emphasize that the bright compact X-ray source region P5 lies well outside the radio emission on high-resolution total intensity maps.

\item High-polarized intensity feature extends to the North from the hotspot, terminating in the moderately prominent spot with some enhanced level of the diffuse X-ray emission; this spot is selected for our analysis as the extended region ``B''.
 
 \item Upstream of the entire double structure of the E hotspot, an another prominent and extended feature can easily be seen on the polarized intensity map, which in addition seems surrounded by an arc of a sharp RM gradient ($\Delta$RM\,$\sim 50$\,rad\,m$^{-2}$). A particularly bright point/compact X-ray source is located at the Southern edge of the feature; this source is selected for our spectral analysis as region ``P1''. Meanwhile, the Northern-Eastern edge of the feature is surrounded by a prominent enhancement of the diffuse X-ray emission, appearing as an elongated filament that runs in between high-polarized radio intensity domains; this filament is selected for our spectral analysis as region ``A''. Note that the filament coincides with the local minimum in the RM distribution ($\sim 20$\,rad\,m$^{-2}$). The radio continuum at the position of the filament appear marginally steeper when compared with its surroundings (spectral index $\sim 0.9$ versus $\sim 0.8$), although such a minor difference seems rather insignificant taking into account the dynamic range artefacts present on the spectral index map (see the top panel in Figure\,\ref{Fig:Eradio}).
 \end{itemize}

\begin{deluxetable}{llcc}[th!]
\tablecaption{Spectral fitting results for the source region A. \label{Tab:EPLApec}}
\tablewidth{0pt}
\tablehead{\colhead{Model$^{\dagger}$} & \colhead{Parameter} & \colhead{Value with $1\,\sigma$ errors} & \colhead{C-stat./DOF}}
\startdata
 Power-law & Photon index $\Gamma$ &	 $1.71_{-0.22}^{+0.24}$ & 	1077.31/888	\\
 & PL normalization &	$4.37_{-0.52}^{+0.56} \times 10^{-6}$	& 		\\
 & Background photon index $\Gamma_{\rm bck} $ & $0.25_{-0.08}^{+0.08}$	& 		\\
 & Background normalization &	$6.44_{-0.53}^{+0.56} \times 10^{-6}$	& 		\\
\hline
  APEC & Temperature $kT$ &	 $8.22_{-3.20}^{+12.18}$ & 	1079.54/888	\\
 & APEC normalization &	$1.78_{-0.20}^{+0.29} \times 10^{-5}$	& 		\\
 & Background photon index $\Gamma_{\rm bck} $ & $0.26_{-0.08}^{+0.08}$	& 		\\
 & Background normalization &	$6.46_{-0.53}^{+0.57} \times 10^{-6}$	& 		\\
 \hline
  Power-law+APEC & Photon index $\Gamma$ &	 $1.27_{-0.41}^{+0.27}$ & 	1074.3/886	\\
 & PL normalization &	$3.12_{-0.92}^{+0.78} \times 10^{-6}$	& 		\\
 & Temperature $kT$ &	 $0.27_{-0.07}^{+0.14}$ & 		\\
 & APEC normalization &	$5.04_{-2.63}^{+4.54} \times 10^{-6}$	& 		\\
 & Background photon index $\Gamma_{\rm bck} $ & $0.27_{-0.08}^{+0.08}$	& 		\\
 & Background normalization &	$6.52_{-0.56}^{+0.61} \times 10^{-6}$	& 		\\
\enddata
\tablecomments{$^{\dagger}$ all the models include the Galactic hydrogen column density $N_{\rm H,\,Gal} = 4.12 \times 10^{20}$\,cm$^{-2}$; thermal model assumes one-third solar abundance.}
\end{deluxetable}

\subsection{Chandra Spectral Analysis}
\label{sec:Espectral}

Due to the limited or even very-limited photon statistics, the ObsID\,14357-extracted, \emph{unbinned} spectra of all the selected source regions specified above and listed in Table\,\ref{Tab:EPL}, were fitted \emph{simultaneously} along with the background within the 0.5--7.0\,keV range, using the C-stat fitting statistics and the Nelder-Mead or Levenberg-Marquardt optimization methods, assuming simple power-law models for each (with Galactic absorption only). The background was chosen as an extended polygon located just outside of the E lobe, and encompassing the E hotspot region, avoiding bright point sources. The results of the fitting are summarized in Table\,\ref{Tab:EPL}.

As follows, the region A appears harder within the {\it Chandra} range when compared with the other extended regions selected for the analysis, in particular with the secondary hotspot C, but the difference in the best-fit photon index ($\Gamma \simeq 1.7\pm0.2$ versus $\simeq 2.2^{+0.6}_{-0.5}$) is not significant statistically. The point/compact sources P1 and P2 are characterized by relatively steep X-ray continua ($\Gamma > 1.8$ within the errors), while the remaining source P3--P5 appear very hard ($\Gamma < 1.5$ within the errors), especially P3, although here the photon statistics is particularly poor. We note that \citet{Hardcastle16} obtained the spectral index $1.76 \pm 0.10$ for the ``whole E hotspot region'', and $1.80 \pm 0.12$ when excluding X1 (=P1) and X3 (=P4) point/point-like sources.

\begin{figure}[th!]
	\centering 
	\includegraphics[width=0.75\textwidth]{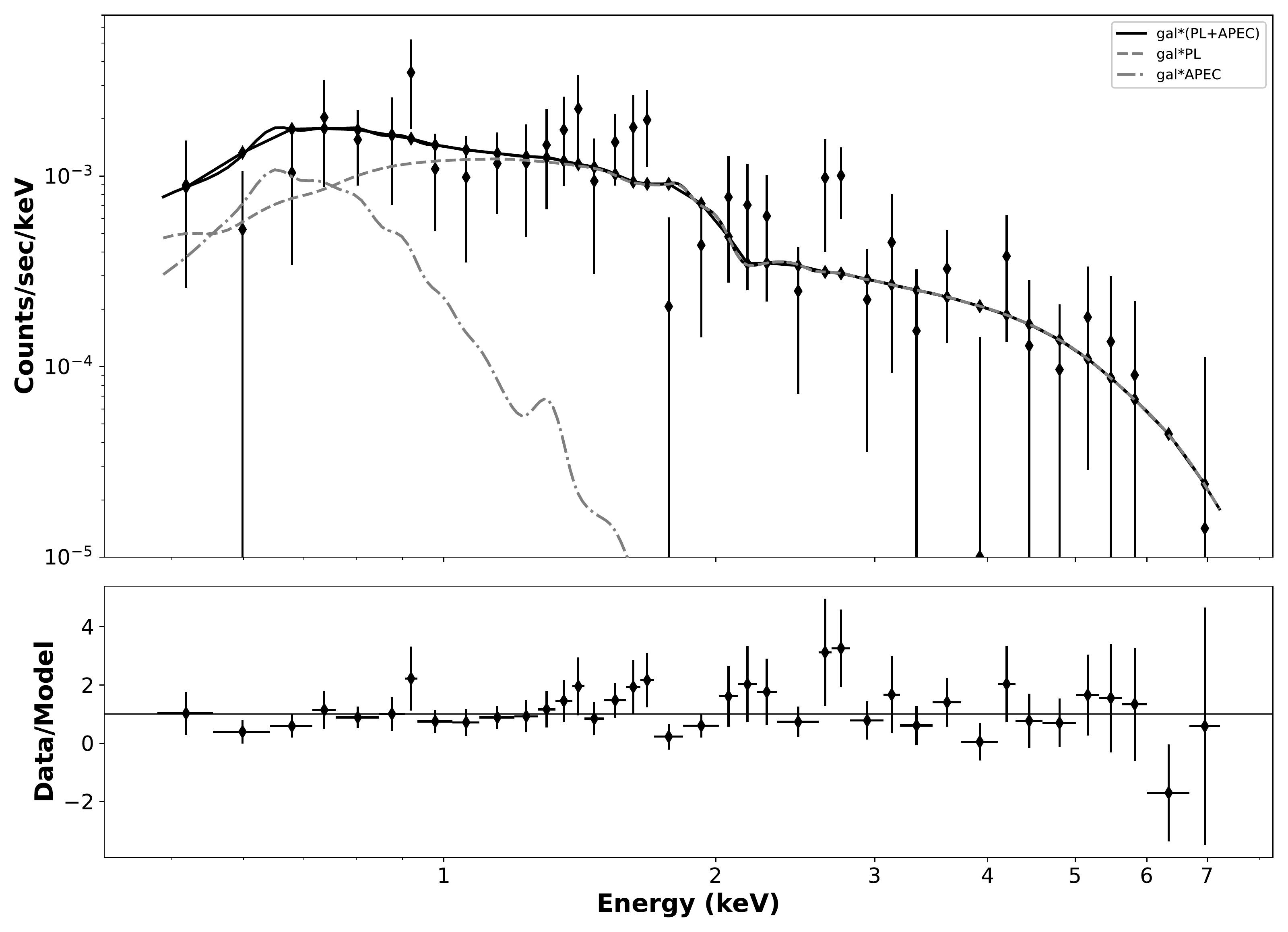}
	\caption{The background-subtracted {\it Chandra} 0.5--7.0\,keV spectrum for the selected source region A, binned with S/N\,$=3$ and fitted with a two-component power-law+APEC model. See Table\,\ref{Tab:EPLApec} for the corresponding best-fit model parameters.}	\label{Fig:ESED}
\end{figure}

The total number of 219 counts detected from the A region, allows us to attempt a more detailed spectral modelling, and in particular to confront the most basic thermal and non-thermal emission models. With this goal in mind, we fit exclusively the ObsID\,14357 spectrum for region A together with the background (same as before), assuming either the power-law emission model, the APEC model, or a combination of the two; in the APEC model, we freeze the abundance at the $0.3$ solar value. The resulting best-fit parameters are summarized in Table\,\ref{Tab:EPLApec}, and the background-subtracted modelled spectrum corresponding to the ``power-law + APEC'' fit, is shown in Figure\,\ref{Fig:ESED}.

\begin{figure}[th!]
	\centering 
	\includegraphics[width=0.95\textwidth]{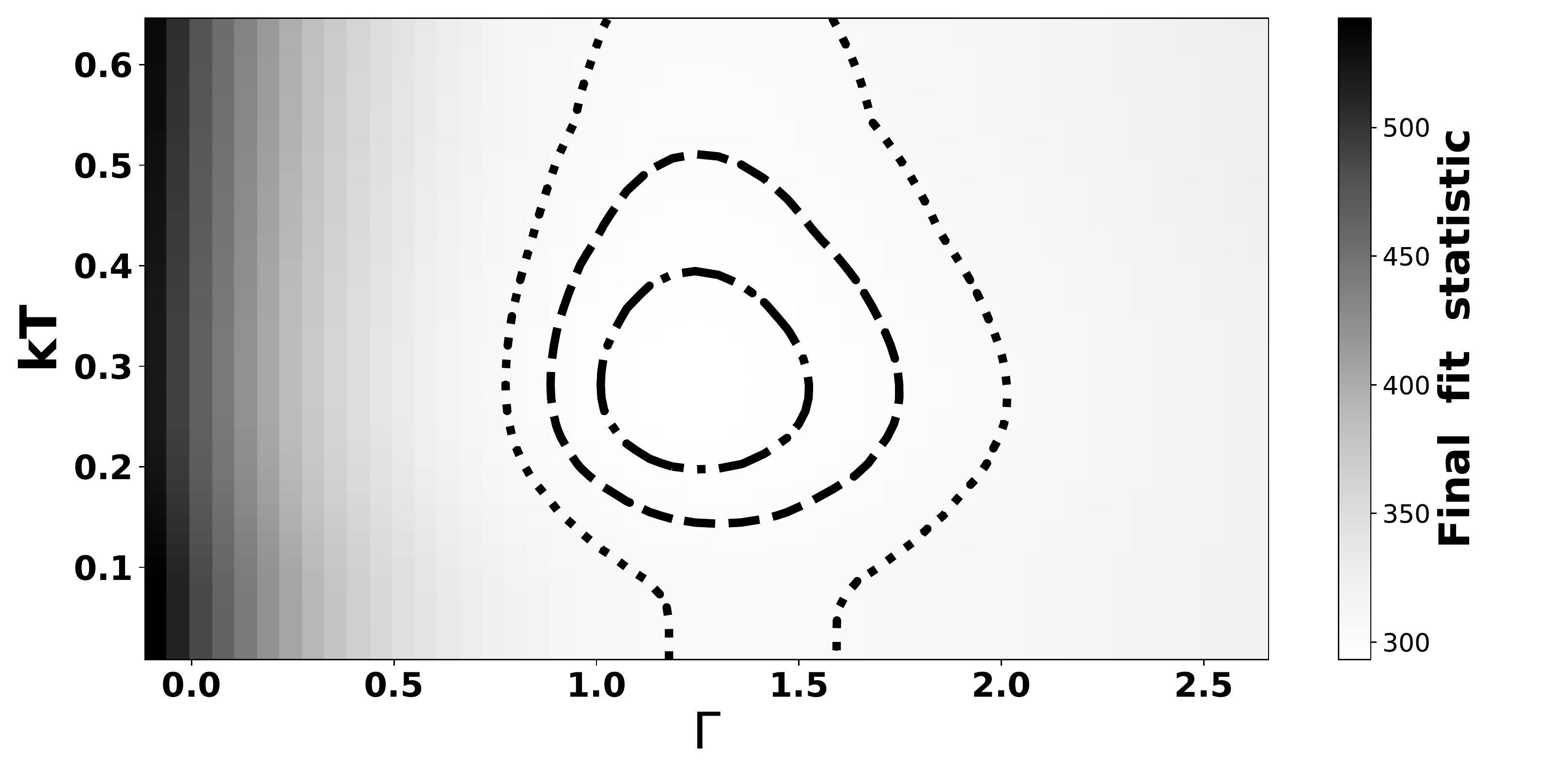}
	\includegraphics[width=0.95\textwidth]{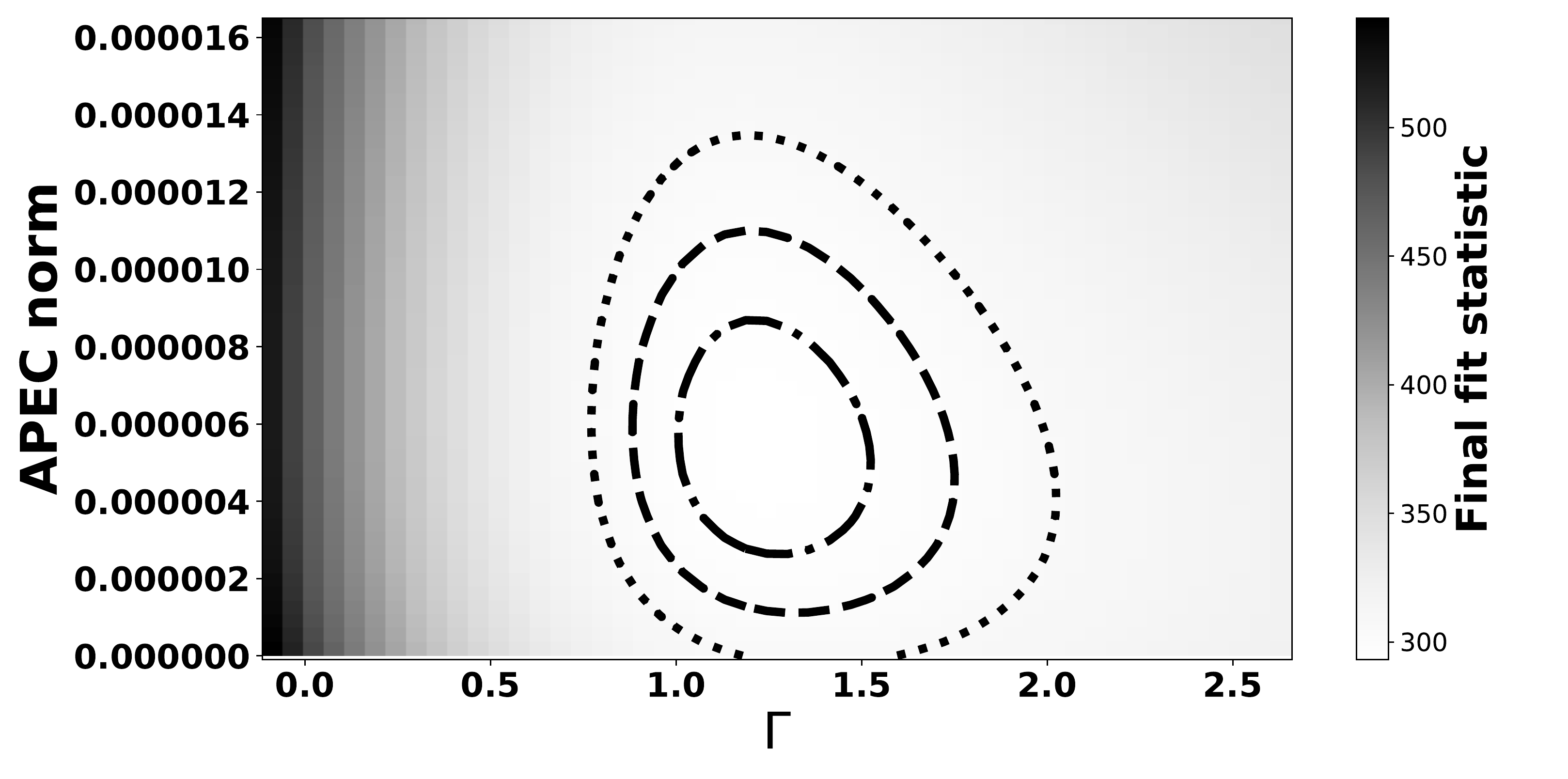}
	\caption{Confidence contours of $1\sigma$, $2\sigma$, and $3\sigma$ on the two-thaw-parameters plane for the ``Power-law + APEC'' model applied to the selected source region A, calculated using the Levenberg–Marquardt optimization method: power-law photon index $\Gamma$ vs. APEC temperature $kT$ (upper panel), and power-law photon index $\Gamma$ vs. APEC normalization (lower panel).}	
	\label{Fig:Econf}
\end{figure}

As follows, the goodness of all the three fits is comparable in terms of the reduced statistics, but the gas temperature in the single APEC model is basically unconstrained, $kT > 5.0$\,keV; for this reason, we do not consider a pure thermal model as plausible. A combination of the power-law and APEC components, on the other hand, returns a rather reasonable gas temperature $kT \simeq 0.3\pm 0.1$\,keV, though implies at the same time a rather flat non-thermal continuum, with $\Gamma \simeq 1.3^{+0.3}_{-0.4}$. The corresponding confidence contours for the model parameters $\Gamma$ and $kT$, as well as $\Gamma$ and the APEC normalization, are presented in Figure\,\ref{Fig:Econf} (upper and lower panels, respectively). All in all, we conclude that, while the presence of a thermal X-ray emitting gas within the analyzed filamentary region A  -- in addition to the non-thermal population of electrons emitting inverse-Compton X-rays --  is allowed by the data, it is not, strictly speaking, required or even favoured by the spectral modelling.

\subsection{The Surface Brightness Profile}
\label{sec:Esurface}

\begin{figure}[ht!]
	\centering 
	\includegraphics[width=0.9\textwidth]{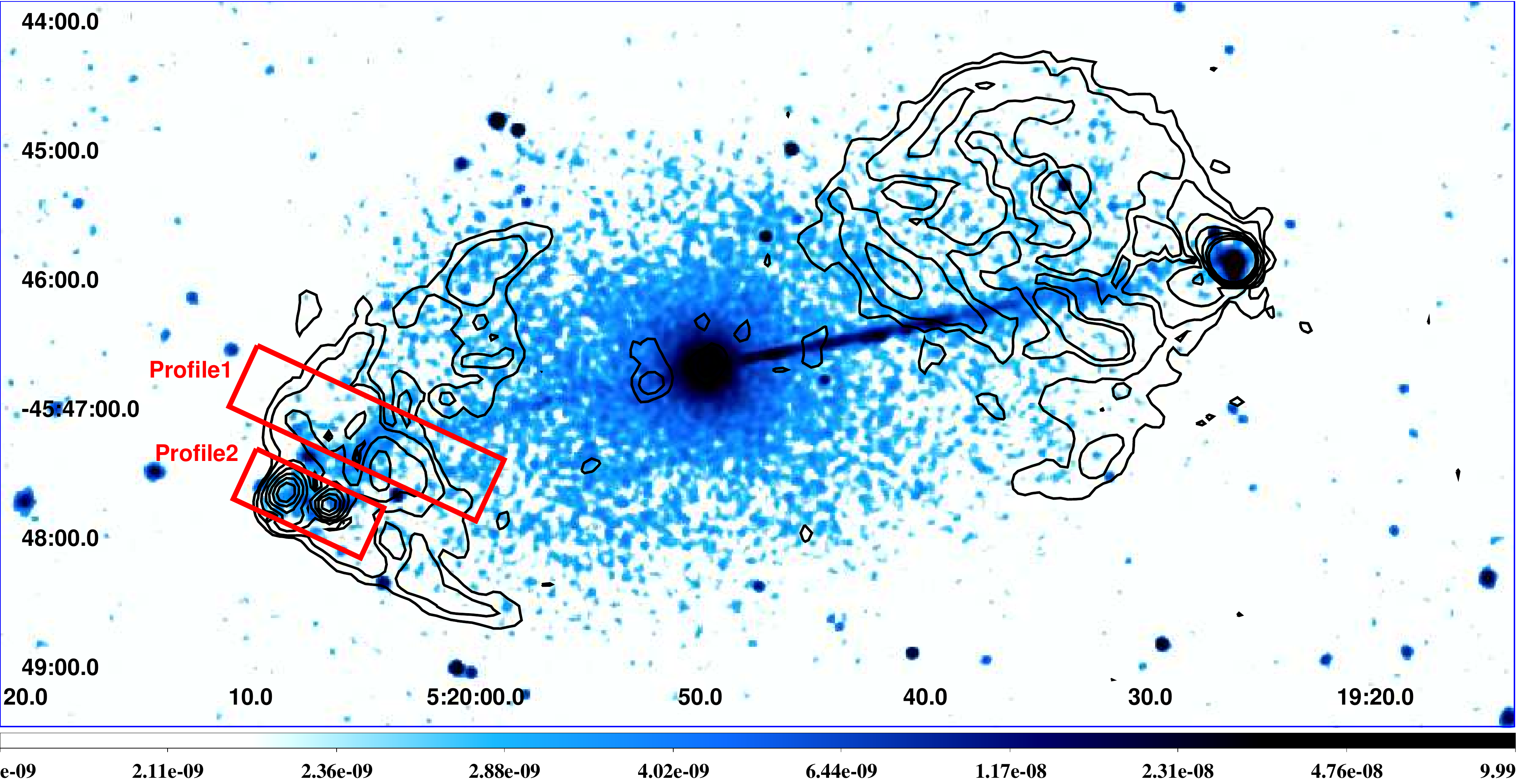}
	\caption{The exposure-corrected 0.5--7.0\,keV merged {\it Chandra} image of the entire structure of Pictor\,A, smoothed with $3\sigma$ Gaussian radius, with the 1.45\,GHz VLA total and polarized intensity ($3\sigma$) contours superimposed (red and white, respectively). The two elongated yellow rectangles, denote the areas across the high-polarization regions of the E lobe, for which we extracted the surface brightness profiles at X-ray and radio frequencies.}
	\label{Fig:Eprof}
\end{figure}

In Figure\,\ref{Fig:Eprof}, we present again the exposure-corrected 0.5--7.0\,keV merged {\it Chandra} image of the entire structure of Pictor\,A, with the 1.45\,GHz VLA total and polarized intensity ($3\sigma$) contours superimposed. The two elongated yellow rectangles, denote the areas across the high-polarization regions of the E lobe, for which we extracted the surface brightness profiles at X-ray and radio frequencies. The `Profile\,2' region includes the double structure of the E hotspot, while the `Profile\,1' region includes the X-ray filament A .

\begin{figure}[ht!]
	\centering 
	\includegraphics[width=0.55\textwidth]{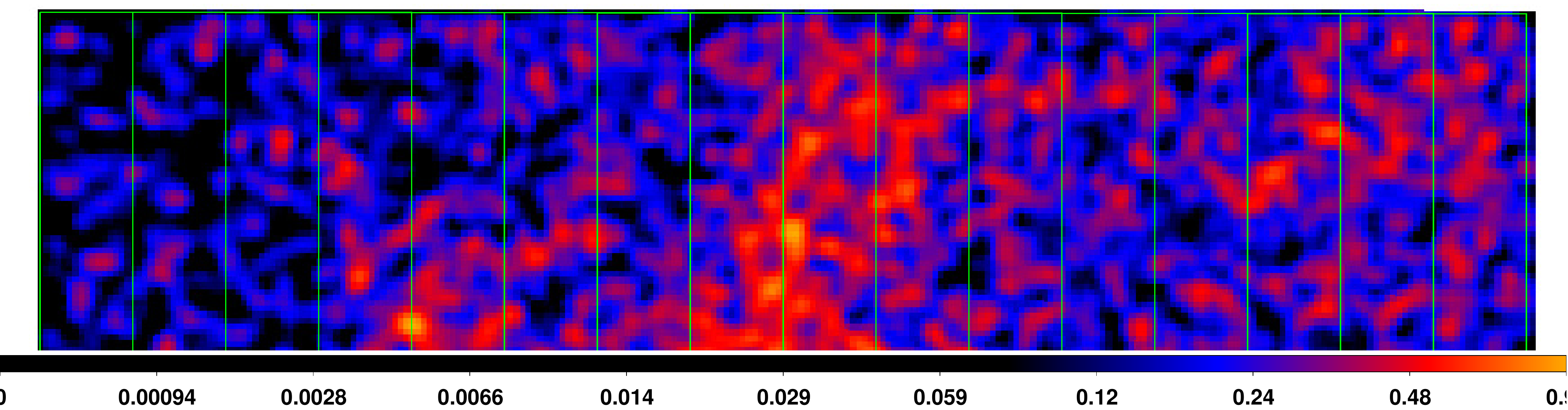}
	\includegraphics[width=0.65\textwidth]{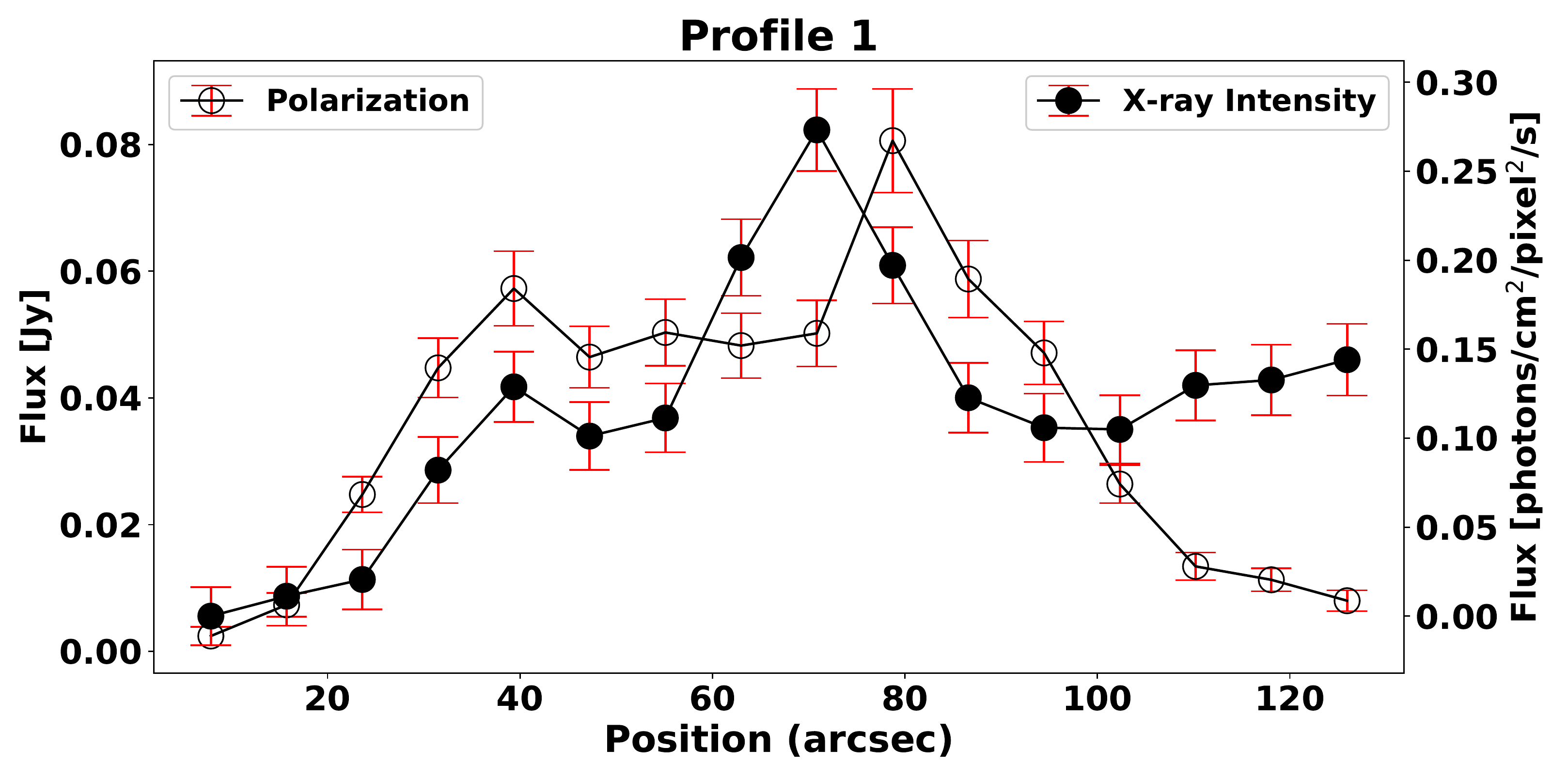}		
	\includegraphics[width=0.65\textwidth]{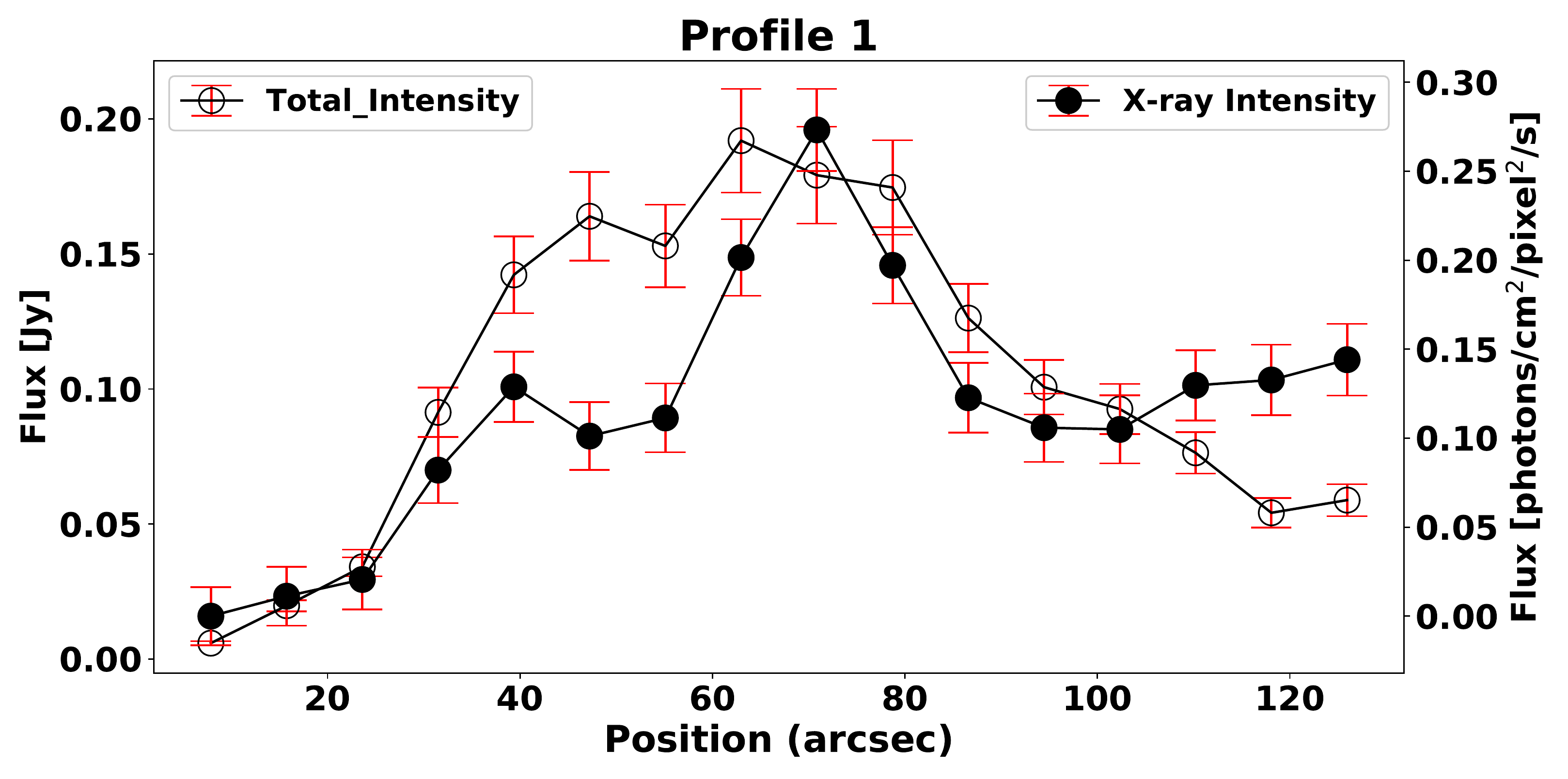}
		\includegraphics[width=0.65\textwidth]{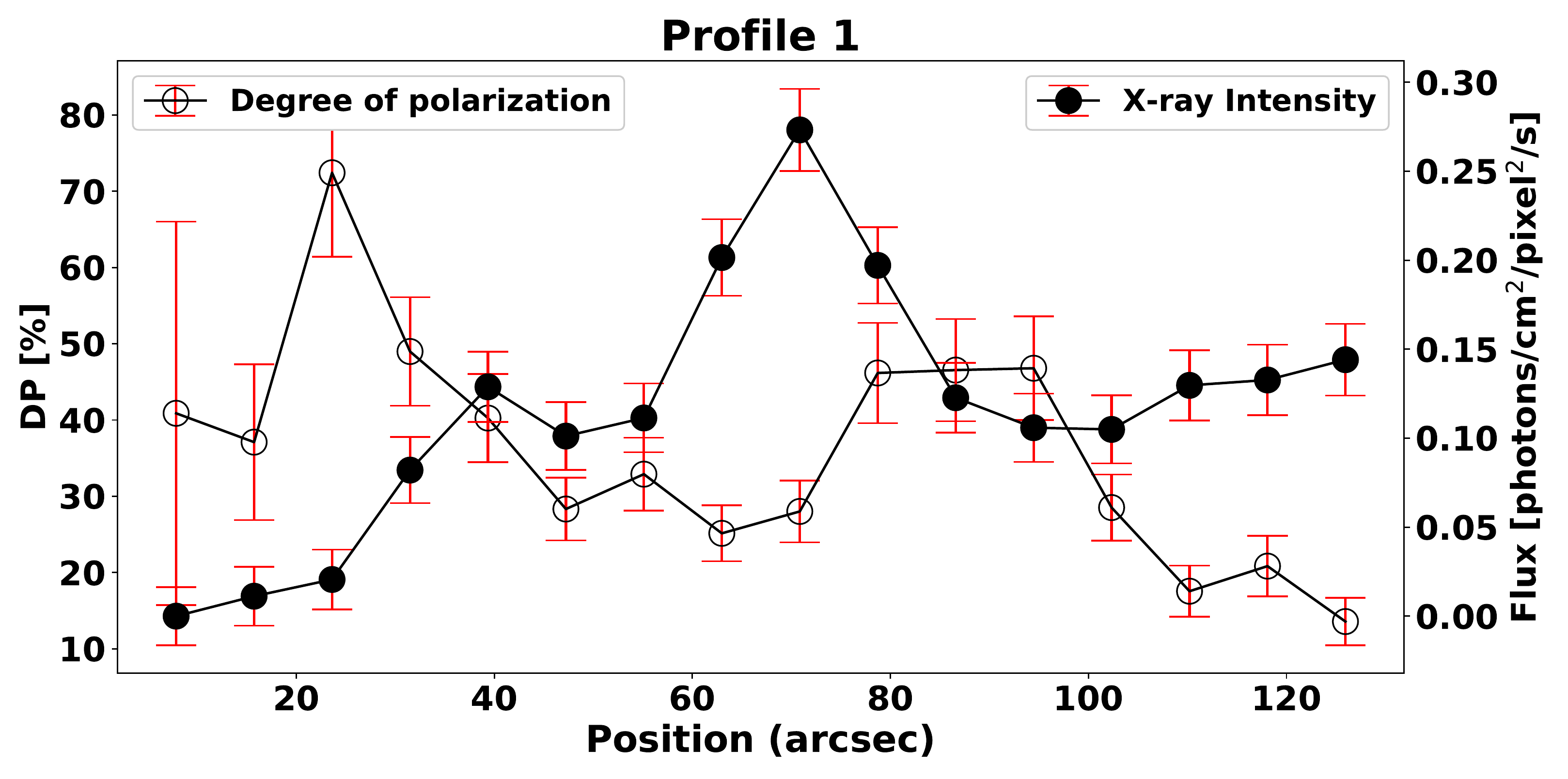}
	\caption{\textbf{Top:} The Profile\,1 rectangular region, rotated by $\theta = 335^{\circ}$, and divided into 16 vertical boxes. \textbf{Middle top:} The X-ray photon fluxes per unit area (filled circles), and the polarized radio flux densities (empty circles), integrated over each box. \textbf{Middle bottom:} The X-ray photon fluxes per unit area (filled circles), and the total radio flux densities (empty circles), integrated over each box. 
	\textbf{Bottom:} The X-ray photon fluxes per unit area (filled circles), and the degree of radio polarization (empty circles), integrated over each box.  }
	\label{Fig:Eprof1}
\end{figure}

\begin{figure}[ht!]
	\centering 
	\includegraphics[width=0.55\textwidth]{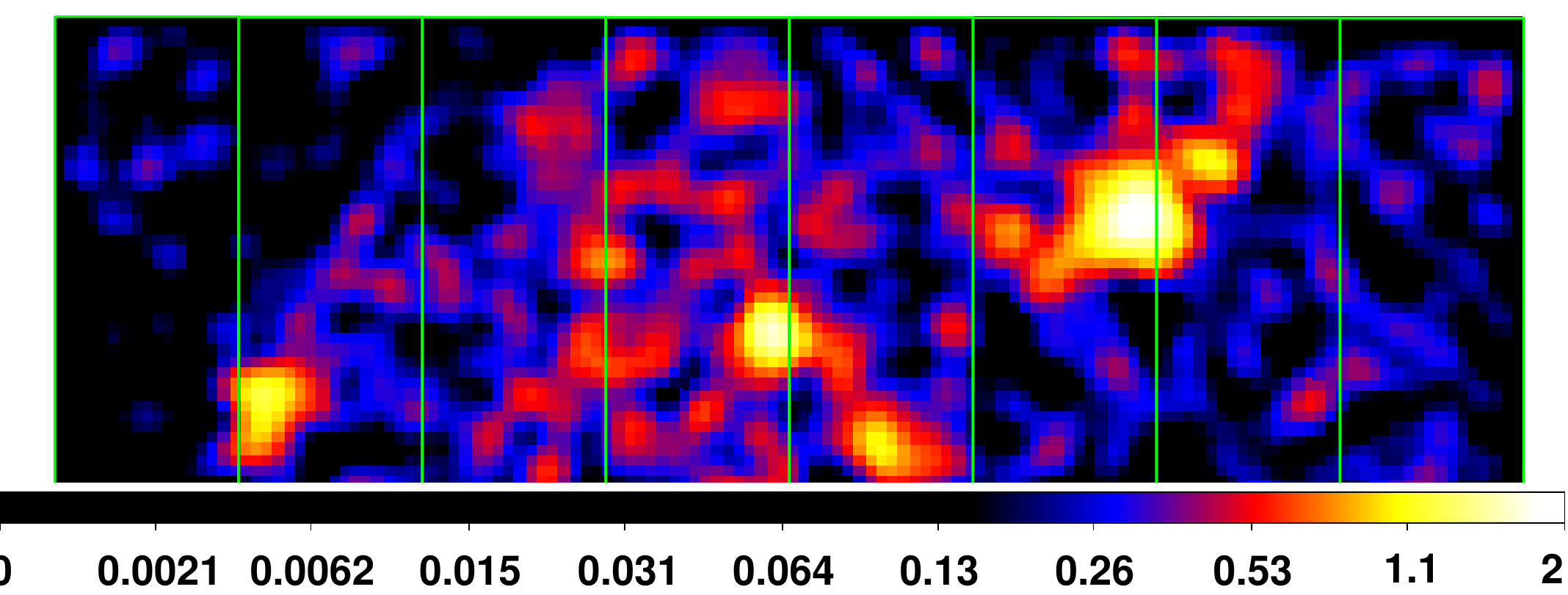}
	\includegraphics[width=0.65\textwidth]{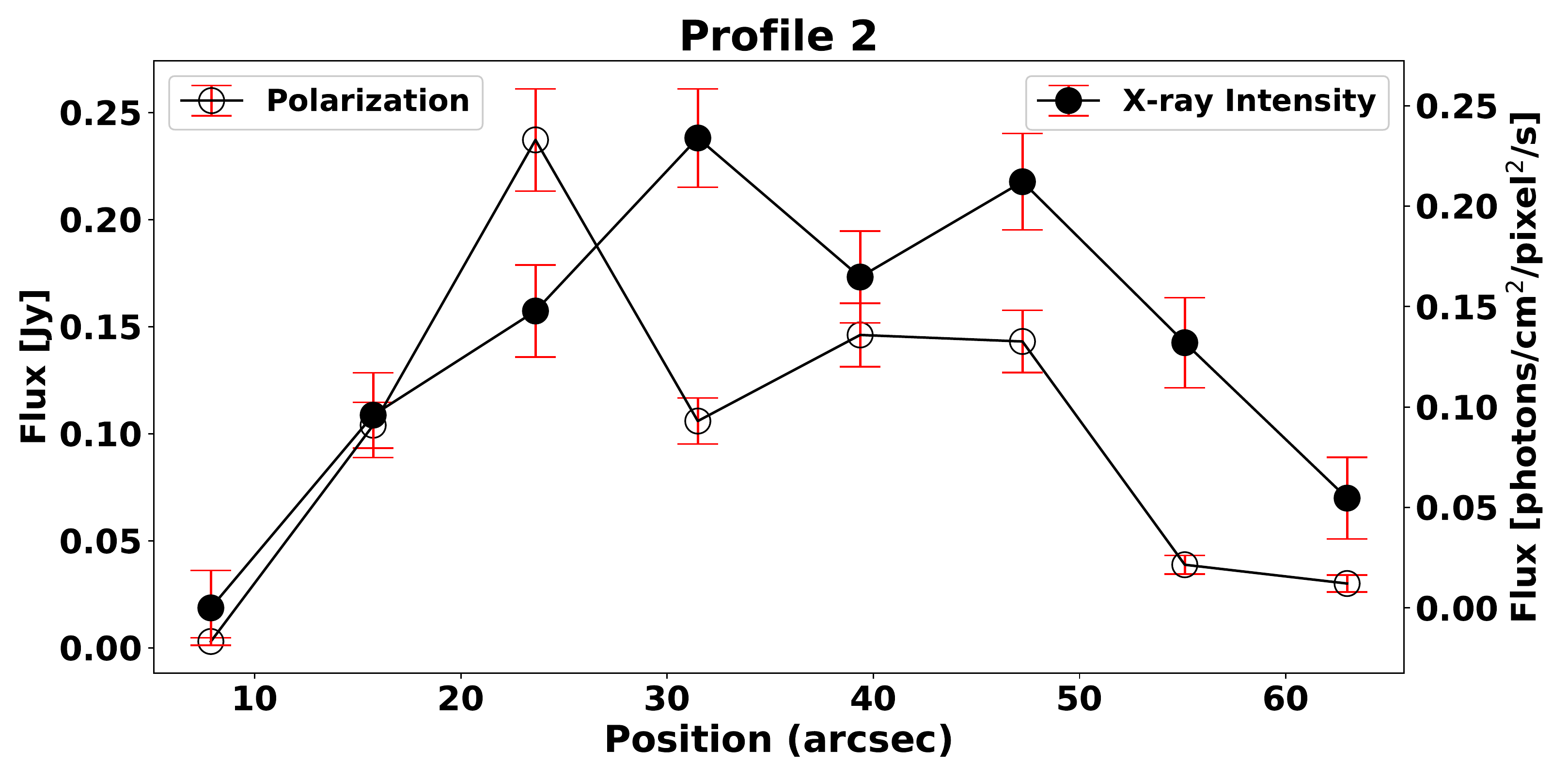}
	\includegraphics[width=0.65\textwidth]{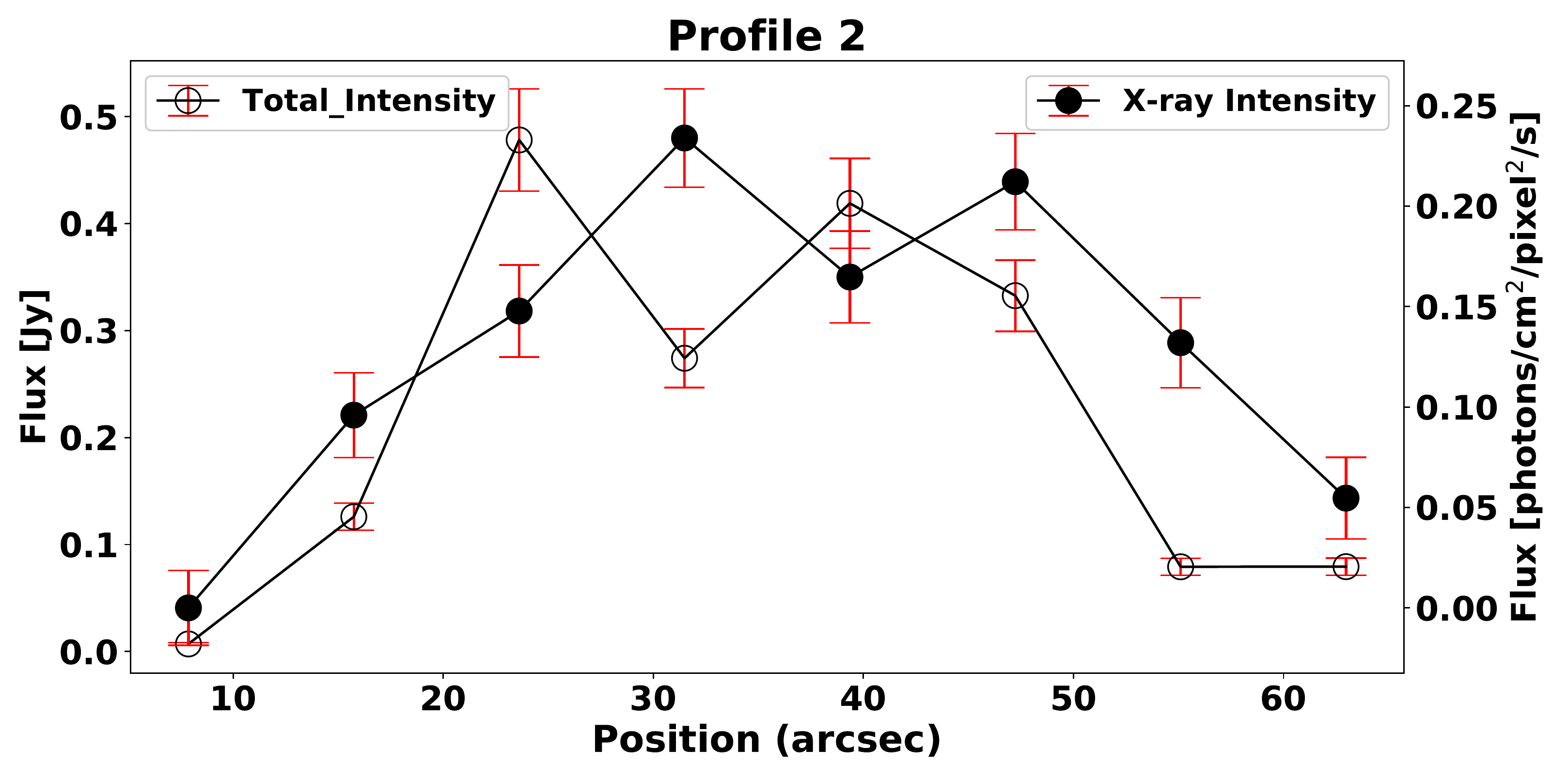}
		\includegraphics[width=0.65\textwidth]{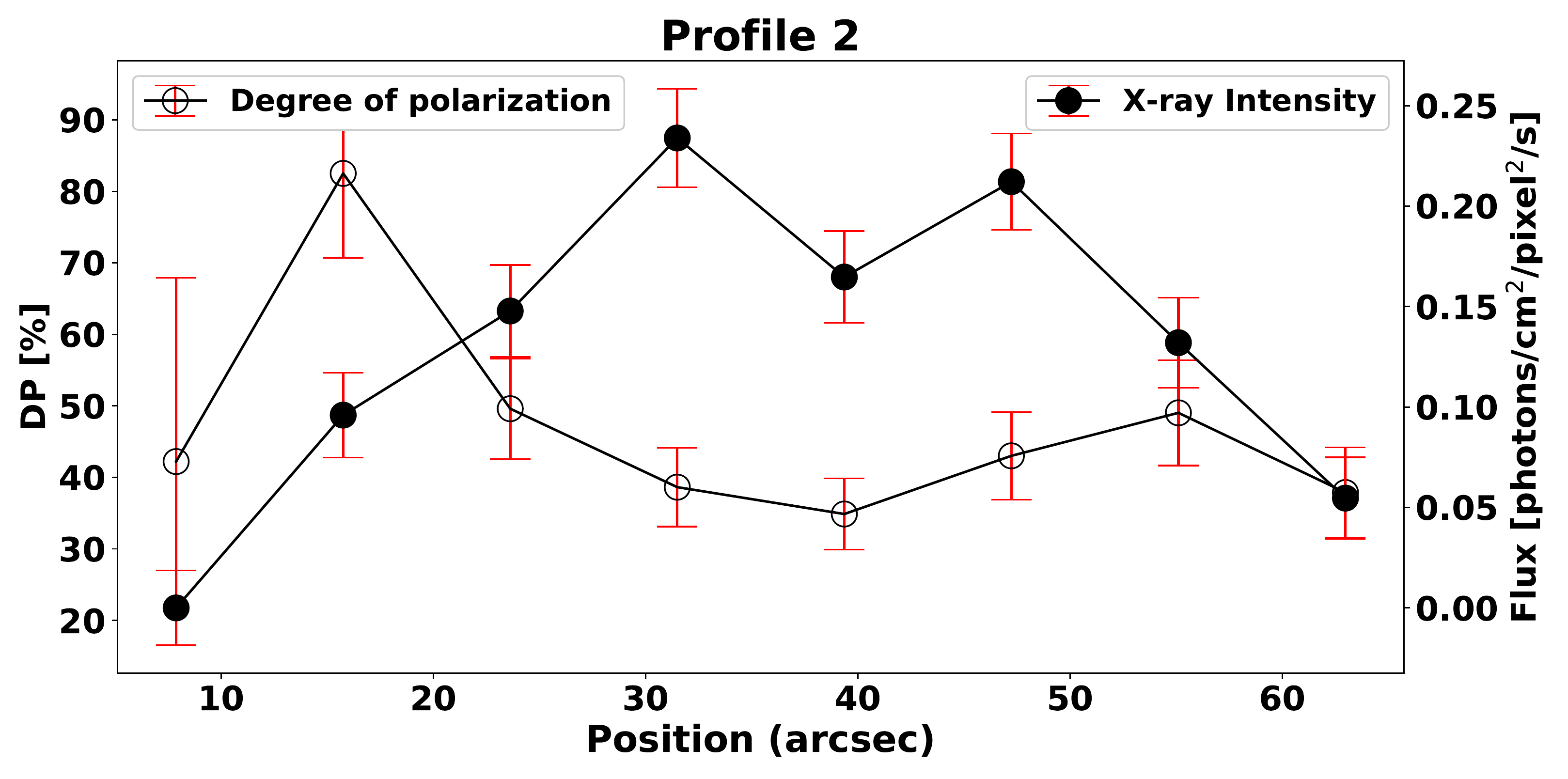}
	\caption{ \textbf{Top:} The Profile\,2 rectangular region, rotated by $\theta = 360^{\circ}$, and divided into eight vertical boxes. \textbf{Middle top:} The X-ray photon fluxes per unit area (filled circles), and the polarized radio flux densities (empty circles), integrated over each box. \textbf{Middle bottom:} The X-ray photon fluxes per unit area (filled circles), and the total radio flux densities (empty circles), integrated over each box. 
	\textbf{Bottom:} The X-ray photon fluxes per unit area (filled circles), and the degree of radio polarization (empty circles), integrated over each box.  }
	\label{Fig:Eprof2}
\end{figure}

The Profile\,1 region, rotated by $\theta = 335^{\circ}$ using {\fontfamily{qcr}\selectfont dmregrid2}, is divided into 16 vertical boxes, as indicated in the upper panel of Figure\,\ref{Fig:Eprof1}. X-ray counts were extracted from the merged {\it Chandra} map within the energy range 0.5--7.0\,keV (binsize=1), summed for each box, and then converted to the surface brightness units as recommended by the CIAO 4.13 Science Threads.\footnote{ \url{https://cxc.cfa.harvard.edu/ciao/threads/radial_profile/}} The total and polarized radio flux densities for the corresponding segments of the lobe, were calculated based on the 1.45\,GHz VLA maps; polarization degree was calculated as a ratio of the polarized and total intensities. The resulting profiles are presented in the three bottom panels of Figure\,\ref{Fig:Eprof1}. In an analogous way, we calculated the X-ray and radio profiles also for the Profile\,2 rectangular region, divided into eight vertical boxes, and rotated by $\theta = 360^{\circ}$, see Figure\,\ref{Fig:Eprof2}.

Figure\,\ref{Fig:Eprof1} reinforces our main observational finding stated above, that the prominent X-ray filament\,A runs in between high-polarization radio domains: while the total radio intensity peak ($\lesssim 0.2$\,Jy) coincides roughly with the integrated X-ray surface brightness maximum, the polarized radio flux peak ($\sim 0.08$\,Jy) is located upstream the X-ray maximum. As a result, the degree of radio linear polarization increases from about $25\%$ at the position of the X-ray filament, up to about $45\%$ at the filament's edges.

The surface brightness profiles shown in Figure\,\ref{Fig:Eprof2} should be taken with caution, due to the fact that the integrated X-ray fluxes here are affected by the presence of the point/compact X-ray sources P2, P3 and P5. Still, we note a much better match between the total and polarized radio fluxes in this case (as compared with Profile 1), with both peak maxima shifted downstream (i.e., to the East) with respect to the X-ray surface brightness maxima; also, we note that the primary and secondary hotspots display in general a very high degree of radio polarizaion, at the level of $\sim 40-50\%$.

\section{Discussion and Conclusions}

A relation of point-like/compact X-ray features with no optical counterparts, to the radio lobes and especially hotspot regions of radio galaxies and radio quasars, is often unclear and subjected to speculations \citep[see the discussion in, e.g.,][]{Hardcastle07}. Such features may simply be unrelated background AGN, but may also result from various energy dissipation processes taking place within the lobes with complex magnetic field structure. For example, \cite{Stawarz13} speculated that, if the lobes' radio filaments represent indeed tangled magnetic field tubes \citep{ONeill10}, then at the places of the filaments' interactions with density or magnetic enhancements in the surrounding plasma, localized multiple compact sites of violent reconnection may form, loading turbulence and in this way enabling efficient particle acceleration and plasma heating. 

Among the point/compact X-ray sources P1--P5 analyzed here, none possesses any obvious optical counterpart. However, the steep-spectrum P1 spot coincides exactly with the mid-infrared (MIR) source listed in the CatWISE2020 Catalog \citep{Marocco20}, which includes objets selected from WISE \citep{Wright10} and NEOWISE \citep{Mainzer14} all-sky survey at 3.4 and 4.6\,$\mu$m (W1 and W2), conducted between 2010 to 2018. In fact, in the same catalog, possible MIR counterparts to P4 and P4 may also be noted (each with the separation below $6\arcsec$). However, no MIR features have so far been detected at and around the positions of P2 and P3, and so we believe that at least those two spots may be plausible candidates for the lobe-related compact X-ray structures. We note in this context, that the two spots are located just upstream of the primary hotspot and the secondary hotspot, respectively. In addition, P3 seems extremely hard in X-rays, as revealed by our spectral modelling despite a very low photon statistics (photon index $\Gamma < 1.2$ within the errors; see Table\,\ref{Tab:EPL}), while P2 seems partly resolved by {\it Chandra} \citep[see][]{Hardcastle16}.

The main findings following from the analysis presented in this paper, regard however the elongated X-ray filament A, located upstream of the jet termination region, extending for at least 30\,kpc (projected), and inclined with respect to the jet axis. Its 0.5--7.0\,keV radiative output is consistent with a pure power-law emission with the photon index $\Gamma \simeq 1.7\pm0.2$, or alternatively a combination of a flat power-law component with $\Gamma \simeq 1.3^{+0.3}_{-0.4}$ and a thermal $kT \simeq 0.3\pm 0.1$\,keV plasma. In the former case, the X-ray slope would be consistent (within the errors) with the slope of the radio continuum at the position of the filament. The latter case would, on the other hand, be in accord with recent findings of a larger amount of a thermal gas within the radio lobes of radio galaxies \citep{Stawarz13,OSullivan13}. 

It is interesting to note in this context a scenario proposed by \cite{Anderson18}, in which the low-polarization regions associated with the magnetic field's line-of-sight reversals, as observed within the radio lobes of the Fornax\,A galaxy, are due to a thermal matter with mass $\mathcal{O}(10^9 M_{\odot})$, distributed within thin shells or filaments. Possibly, the observed characteristics of the X-ray filament A in the E lobe of Pictor\,A, as presented in this paper, conform to this particular model. The degree of the radio linear polarization does increase from about $25\%$ at the filament's axis, up to about $45\%$ at the filament's edges, and this could indeed be due to the depolarization related to the thermal (X-ray emitting) gas present within the filament. But at the same time the filament coincides also with the local minimum in the RM distribution, and this would then imply that the magnetic field increases its net out-of-line-of-sight component, at the expense of the net line-of-sight component $B_{\parallel}$, since RM\,$\propto \int^L n_e \, \vec{B}\cdot d\vec{\ell}$\,$\sim 0.8 \, (n_e/{\rm 10^{-3}\,cm^{-3}}) \, (B_{\parallel}/{\rm \mu G}) \, (L/{\rm kpc})$\,rad\,m$^{-2}$, where $n_e$ stands for the gas electron number density, $\vec{B}$ is the magnetic field intensity vector, and the integration $d\vec{\ell}$ is over the path length through the plasma.

For a rough estimate in this context, let us approximate the X-ray filament A by an elongated ellipsoid with the major axis $a \simeq 80$\,px\,$\sim 27.5$\,kpc, and the minor axis $b \simeq 48$\,px\,$\sim 16.5$\,kpc. For such, the volume reads as $V = \frac{4}{3}\pi (a/2) (b/2)^2 \sim 10^{68}$\,cm$^{3}$, and --- given the APEC normalization provided in Table\,\ref{Tab:EPLApec} --- the thermal gas electron density $n_e \sim [5 \times 10^{-6} \, 4 \pi d_{\rm L}^2 / 10^{-14} (1+z)^2 \, \mu V]^{1/2} \sim 3.8 \times 10^{-3}$\,cm$^{-3}$, assuming uniform ionized plasma with the ratio of proton-to-electron number densities $\mu \equiv n_{\rm H}/n_e \simeq 0.82$. This corresponds to the total mass in the filament $M \simeq m_p \, \mu n_e \, V \sim 3 \times 10^8 M_{\odot}$, which is in a very good agreement with the model by \citet{Anderson18}.  Moreover, the total pressure of the thermal gas, $p_g = n_e \, kT \simeq 1.6 \times 10^{-12}$\,dyn\,cm$^{-2}$, would be then comparable to that of the lobes' magnetic field for the magnetic field intensity $B\simeq 6$\,$\mu$G, which is, in fact, very close to the equipartition value, i.e., the value obtained by assuming pressure balance between ultra-relativistic radio-emitting electrons and the lobes' magnetic field. We note in this context that, based on the inverse-Compton modeling of the observed X-ray emission, \citet{Hardcastle16} have estimated the mean magnetic field strength in the extended lobes of Pictor\,A as $\langle B \rangle \simeq 4$\,$\mu$G, which is a factor of 1.5 below the equipartition level.

With the above estimates, one can show that the depolarization effect due to the RM dispersion in the filament, may be quite substantial indeed. For this, first we note that the depolarization of the (background) lobes' emission at a wavelength $\lambda$, by an external Faraday screen (i.e., the embedded thermal filament), is given by
\begin{equation}
{\rm DP}(\lambda) \propto \exp\left[- 2 \sigma_{\rm RM}^2 \lambda^4\right] \, ,
\end{equation}
where $\sigma_{\rm RM}$ is the standard deviation of the RM within the beam, related to the tangled magnetic field \citep{Burn66,Laing84}. The  $\sigma_{\rm RM}$ parameter can be estimated by assuming that the Faraday screen consists of cells of coherent magnetic field with the size scale $d$, and that all the dispersion comes from the line-of-sight field reversals, so that the product of the gas density and $B_{\parallel}$ is roughly equal to $n_e \times \langle B \rangle$ \citep[see][]{Breugel84,Knuettel19}. This gives us
\begin{equation}
\sigma_{\rm RM} \simeq 0.081 \, \left(\frac{n_e}{{\rm cm^{-3}}}\right) \, \left(\frac{\langle B \rangle}{{\rm \mu G}}\right) \, \left(\frac{d}{{\rm kpc}}\right)^{1/2} \, \left(\frac{R}{{\rm kpc}}\right)^{1/2} \, {\rm rad \, cm^{-2}} \, ,
\end{equation}
where $R$ is the depth of the screen. Hence, taking $n_e \simeq 3.8 \times 10^{-3}$\,cm$^{-3}$, $\langle B \rangle \simeq 4$\,$\mu$G, as well as --- for illustrative purposes --- $R \sim 16.5$\,kpc (i.e., the maximum size of the screen) and $d = 0.7$\,kpc (i.e., $10\%$ of the resolution), we obtain $2 \sigma_{\rm RM}^2 \lambda^4 \sim 7$ at $\lambda = 21$\,cm, and $\sim 0.05$ at $\lambda = 6$\,cm, meaning that the 21\,cm lobes' radio emission may be substantially de-polarized by the thermal magnetized gas within the filament.

Clearly, a thermal model for the X-ray filaments within extended lobes of radio galaxies, is not the only plausible scenario. For example, keeping in mind that such lobes may be acceleration sites of ultra-high energy cosmic rays \citep[UHECRs; see the discussion in, e.g.,][]{Hardcastle09,OSullivan09}, one could speculate following \citet{Stawarz13} that the flux of UHECRs through the lobes may produce uncompensated currents, which locally alter both the magnetic field configuration and the electron heating/acceleration conditions, via cosmic ray current-driven instability. In this model, it would however remain rather unclear why, an enhancement in the non-thermal electron population, manifesting as an enhancement in the X-ray surface brightness via inverse-Comptonization of the Cosmic Microwave Background photons, should be associated with a lower radio polarization, and/or with a decrease in the RM value. This demonstrates that a proper discrimination between various models and plausible scenarios proposed for the internal structure of the extended lobes in radio galaxies and quasars, requires not only a deep X-ray imaging of the lobes on arcsec-scales, but also the accompanying detailed radio polarimetry with a comparable high resolution.

\begin{acknowledgements}
This research has made use of data obtained from the Chandra Data Archive. This work was supported by the Polish NSC grant 2016/22/E/ST9/00061. The authors thank the anonymous Referee, as well as K. Nalewajko, for all their relevant comments and suggestions, which helped to improve the paper.
\end{acknowledgements}

\vspace{5mm}
\facilities{Chandra(ACIS)}

\software{CIAO \citep{Fruscione06}, Sherpa \citep{Freeman01}}

\end{document}